\documentclass[%
reprint,
superscriptaddress,
 amsmath,amssymb,
 aps,
 prl,
 twocolumn
]{revtex4-1}

\usepackage{graphicx}
\usepackage{dcolumn}
\usepackage{bm}
\usepackage{xr}
 \usepackage[colorlinks]{hyperref}

\usepackage{physics}
\usepackage{multirow}
\usepackage{makecell}
\usepackage{array}

\newcommand{\da}{\dagger}

\newcommand{\hc}{\text{h.c.}}

\newcommand{\dr}{\text{dr}}

\newcommand{\tar}{\text{tar}}

\newcommand{\tot}{\text{tot}}

\newcommand{\opt}{\text{opt}}

\newcommand{\inp}{\text{in}}
\newcommand{\at}{\tilde{\alpha}}

\newcommand{\dbp}{\text{dp}}

\usepackage{xcolor}
\usepackage{commath}

\begin{document}

\title{Universal Control in Bosonic Systems with Weak Kerr Nonlinearities}

\author{Ming Yuan}
\email{yuanming@uchicago.edu}
\affiliation{Pritzker School of Molecular Engineering, The University of Chicago, Chicago, IL 60637, USA}
\author{Alireza Seif}
\affiliation{Pritzker School of Molecular Engineering, The University of Chicago, Chicago, IL 60637, USA}
\author{Andrew Lingenfelter}
\affiliation{Department of Physics, The University of Chicago, Chicago, IL 60637, USA}

\author{David I. Schuster}
\affiliation{Pritzker School of Molecular Engineering, The University of Chicago, Chicago, IL 60637, USA}
\affiliation{Department of Physics, The University of Chicago, Chicago, IL 60637, USA}
\affiliation{James Franck Institute, The University of Chicago, Chicago, IL 60637, USA}

\author{Aashish A. Clerk}
\affiliation{Pritzker School of Molecular Engineering, The University of Chicago, Chicago, IL 60637, USA}
\author{Liang Jiang}
\email{liangjiang@uchicago.edu}
\affiliation{Pritzker School of Molecular Engineering, The University of Chicago, Chicago, IL 60637, USA}

\date{\today}

\begin{abstract}
    Resonators with weak single-photon self-Kerr nonlinearities can theoretically be used to prepare Fock states in the presence of a loss much larger than their nonlinearities. Two necessary ingredients are large displacements and a two-photon (parametric) drive. Here, we find that these systems can be controlled to achieve any desired gate operation in 
    a finite dimensional subspace (whose dimensionality can be chosen at will).   Moreover, we show that the two-photon driving requirement can be relaxed and that full controllability is achievable with only 1-photon (linear) drives. We make use of both Trotter-Suzuki decompositions and gradient-based optimization to find control pulses for a desired gate, which reduces the computational overhead by using a small blockaded subspace. We also discuss the infidelity arising from input power limitations in realistic settings, as well as from corrections to the rotating-wave approximation. Our universal control protocol opens the possibility for quantum information processing using a wide range of lossy systems with weak nonlinearities.
    \end{abstract}


\maketitle



\textit{Introduction}--
Bosonic systems, such as photons in optical or microwave resonators, are a promising platform for quantum information processing. In contrast to qubits, the 
infinite-dimensional bosonic Hilbert space provides novel ways to encode and robustly process quantum information in a hardware-efficient manner~\cite{joshi2021quantum}. 
A challenge however is the need for nonlinear operations~\cite{kok_linear_2007}. It has been shown that, to achieve universal control in bosonic systems it is necessary and sufficient to have at least one kind of nonlinear operation in addition to linear operations, i.e., unitary evolution under a Hamiltonian linear or quadratic in bosonic raising and lowering operators~\cite{Lloyd1999, eriksson2023universal}. One approach to introduce nonlinearity is to couple the bosonic system directly to a nonlinear system such as a qubit~\cite{Krastanov2015,Hacohen-Gourgy2016,Eickbusch2022, diringer_conditional_2023}, but for many platforms, achieving this with sufficiently strong coupling can be difficult. For example, in millimeter wave regime (around 100 GHz) there is no good superconducting qubit yet; other bosonic systems like phonons can work in a higher temperature beyond the requirement for superconducting qubits~\cite{chan_laser_2011,patel_room-temperature_2021}.

Given this, it would be ideal to exploit intrinsic nonlinearities in optical or microwave resonators.
An extremely common example is a self-Kerr nonlinearity. Examples include micro-ring resonators or photonic crystals with $\chi^{(3)}$ nonlinearities~\cite{Vernon2015,Choi2017} and quantum LC circuits that contain superconducting materials with high kinetic inductance~\cite{Anferov2020,Xu2023}. While in principle these nonlinearities are sufficient for universal control, in practice they are often much weaker than photon loss rates, precluding the ability to achieve high-fidelity nonlinear operations. 




While nonlinearities weaker than loss might seem to be of no use for quantum nonlinear operations, recent research suggests that this might not be the case~\cite{Lingenfelter2021}. In this work, the authors propose an intriguing scheme to 
deterministically prepare a single-photon Fock state using a Kerr nonlinearity, even in cases where this nonlinearity is significantly smaller than the loss rates.  This scheme relies on a novel photon blockade phenomenon that requires the displacement of a single bosonic mode together with carefully chosen 1-photon and 2-photon driving amplitudes. Notably, the speed of the operations in the blockade subspace is enhanced by the displacement amplitude, which can be large enough to counteract the effects of loss. A similar idea related to displacement-boosted gates is also presented in Ref.~\cite{Eickbusch2022}.


\begin{figure}[t!]
\centering
\includegraphics[scale = 0.7]{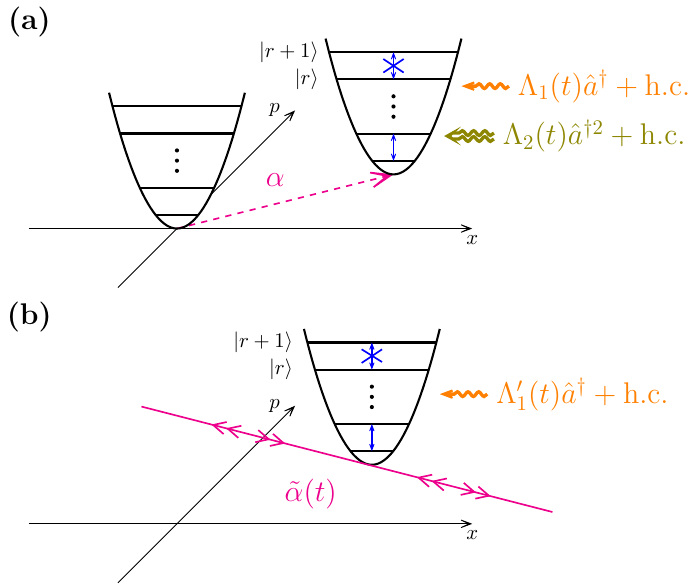}
\caption{Schematic diagram of the photon blockade in the displaced rotating frame with (a) both 1-photon and 2-photon drives, (b) only 1-photon drive with fast oscillation of the displaced rotating frame in phase space.}
    \label{fig:Schemetic}
\end{figure}

Here, we generalize this blockade scheme to demonstrate its applications beyond the preparation of single-photon Fock states. We show that using this scheme, one can perform any unitary operation in a blockaded subspace of Fock states with an arbitrarily chosen dimension. In addition to formal proof in the ideal scenario, we present a gradient-based optimization algorithm to explicitly find the control pulse sequences to implement a desired unitary operation in the blockaded subspace.

In practice, directly implementing the required 2-photon drive can be a challenging task,  as this can require additional weak nonlinearities. For example, in certain platforms one could pump an auxiliary mode that interacts nonlinearly with the central mode~\cite{Bruch2019}.
Here we show that one can eliminate the need for an explicit 2-photon drive, by instead simply time-modulating the amplitude of the single-photon drive.  This is similar in spirit to the operation of double-pumped parametric amplifiers~\cite{Kamal2009}. 

The optimization algorithm and this modulation scheme are then integrated seamlessly to implement arbitrary operations 
using only a standard linear, single-photon drive.  
This simplification comes at the cost of requiring larger 1-photon driving amplitudes, which may result in much stronger input power or even the violation of the rotating wave approximation (RWA). We discuss these implications and possible methods to resolve the resulting imperfections. We argue that even with these limitations, our scheme performs well in experimental platforms where the strength of the single-photon self-Kerr nonlinearity is comparable to or slightly smaller than the loss rate, which was not possible using previous techniques. 

\textit{System setup}--
We consider a setup similar to Ref.~\cite{Lingenfelter2021}, consisting of a single-mode resonator with self-Kerr nonlinearity, subject to both 1-photon and 2-photon drives. Within the RWA, the Hamiltonian of the system can be written as  
\begin{equation}\label{eqn:ham_lab}
\begin{split}
    \hat H = & \frac{\chi}{2} \hat a^{\da 2} \hat a^2 + \omega_c \hat a^\da \hat a + [\Lambda_1(t) e^{-i\omega_1(t) t} \hat a^\da\\
    & + \Lambda_2(t) e^{-i\omega_2(t) t} \hat a^{\da 2} + \hc],
\end{split}
\end{equation}
where $\chi$ indicates the strength of the single-photon Kerr nonlinearity, $\omega_c$ is the angular frequency of the resonator, and $\Lambda_{1(2)}$ and  $\omega_{1(2)}$ are the amplitude and the frequency of the 1(2)-photon drives, respectively. Here, h.c. denotes the Hermitian conjugate of the terms in the bracket. As shown in Ref.~\cite{Lingenfelter2021}, by choosing $\omega_2 = 2\omega_1$, going to a frame rotating with $\omega_1(t)$ and then displaced by $\alpha(t)$, we can obtain a blockaded Hamiltonian of the form
\begin{equation}\label{Hdr}
    \hat H_{\dr}[\alpha(t)] = \frac{\chi}{2} \hat a^{\da 2} \hat a^2 + \Delta_0 \hat a^\da \hat a + [\chi\alpha(t) \hat a^\da (\hat n - r) + \hc],
\end{equation}
where $r$ is an adjustable positive integer determining the block subspace's dimension, and $\Delta_0$ is the detuning term in the displaced rotating frame. To achieve this effective Hamiltonian for a given $\alpha(t)$, $\Delta_0$ and integer $r$, one needs time-dependent drive amplitudes and frequencies chosen such that 
\begin{equation}\label{omega1}
\begin{cases}
\Lambda_1(t) = \chi\alpha(t)[2|\alpha(t)|^2 - r] - \Delta_0 \alpha(t) + i\dot \alpha(t),\\
\Lambda_2(t) = -\frac{\chi}{2}\alpha^2(t),\\
\omega_1(t) = \omega_c - \Delta_0 + \frac{1}{t} \int^t_0 2\chi|\alpha(t')|^2 \dif t'.
\end{cases}
\end{equation}

The key term in Eq.~\eqref{Hdr} is the nonlinear drive, which has no coupling between $\ket{r}$ and $\ket{r+1}$ levels. Therefore, the dynamics generated by $\hat H_\dr$ is constrained within the blockade subspace $\mathcal{H}_b$ spanned by $\{\ket{0}, \ket{1}, \dots, \ket{r}\}$. Also, the amplitude of this nonlinear drive is set by the time-dependent displacement $\alpha(t)$, which serves as a key control parameter in what follows. Previously, it was shown that with a static $\alpha$, one could use $\hat H_\dr$ to generate Fock $\ket{1}$ state on a timescale much shorter than $1/\chi$~\cite{Lingenfelter2021}. Here, we go much further: we show that in fact, by using a time-dependent $\alpha(t)$, one can generate any unitary within the blockade subspace $\mathcal{H}_b$, whose dimension can also be chosen freely.

To account for the effects of photon loss in our system, we use the master equation
\begin{equation}
    \frac{\dif \hat \rho}{\dif t} = -i [\hat H, \hat \rho] + \kappa\mathcal{D}[\hat a]\hat \rho,
\end{equation}
where $\mathcal{D}[\hat{a}] \hat\rho = \hat{a} \hat{\rho} \hat{a}^\dagger - \frac{1}{2}\{\hat{a}^\dagger \hat{a},\hat{\rho}\}$ is a dissipator that models the photon loss effect and $\kappa$ is the rate of this process. Note that when transforming to the same displaced rotating frame mentioned earlier, the 1-photon driving amplitude $\Lambda_1(t)$ in Eq.~\eqref{omega1} now requires the addition of an extra term, namely $i\kappa\alpha(t)/2$, in order to obtain the same Hamiltonian $\hat H_\dr[\alpha(t)]$.  However, the form of the dissipator $\mathcal{D}[\hat a]$ and the associated loss rate $\kappa$ remain unaltered in this new frame.
Consequently, due to the enhancement of the nonlinear blockade drive by $\alpha$ in $\hat H_\dr$, the operations can be performed on a significantly shorter timescale than $1/\chi$. This, in turn, allows for the mitigation of the impact of photon loss, presenting an opportunity to achieve high-fidelity gates, even when $\chi \ll \kappa$.

\textit{Demonstration of universality}--
We sketch the proof of universal controllability of our system governed by the Hamiltonian $\hat H_\dr$ shown in Eq.~\eqref{Hdr} here and refer the reader to Ref.~\cite{supp} for more details. We first focus on the dynamics within the blockade subspace $\mathcal{H}_b$. Let $\hat \Pi_r := \sum_{n=0}^{r} \ketbra{n}$ denote the projector to this $N=r+1$ dimensional subspace. The projection of $\hat H_\dr$ to the blockade subspace is given by
\begin{equation}\label{eq:ham_proj}
    \hat\Pi_r \hat H_\dr \hat\Pi_r = \hat H_{d,0} + \chi\Re[\alpha(t)] \hat H_{c,R} + \chi\Im[\alpha(t)] H_{c,I},
\end{equation}
where
\begin{equation}\label{eq:ham_proj_detail}
\begin{cases}
\hat H_{d,0}=\sum_{n = 0}^r \left[\chi(n^2 - n)/2 + \Delta_0 n\right] \ketbra{n}{n},\\
\hat H_{c,R}=\sum_{n = 0}^{r-1} (n-r)\sqrt{n+1}(\ketbra{n+1}{n} + \hc),\\
\hat H_{c,I}=i\sum_{n = 0}^{r-1} (n-r)\sqrt{n+1}(\ketbra{n+1}{n} - \hc).
\end{cases}
\end{equation}
In what follows (using the language of quantum control theory, see e.g.,~\cite{boscain2021intro}) $\hat H_{d,0}$ serves as the drift Hamiltonian, while $\hat H_{c,R}$ and $\hat H_{c,I}$ are the control Hamiltonians.  The real and imaginary parts of $\alpha(t)$ are time-dependent functions that can be controlled. 
Following Ref.~\cite{Krastanov2015} we define universal control of a quantum system as the ability to realize any unitary operation $\hat U_\tar$ in the U($N$) group via a properly chosen control ($\alpha(t)$ in this case) and evolution time $T$. 

A theorem in Ref.~\cite{Schirmer2001} suggests a sufficient condition for the Hamiltonian to make the system universally controllable. It has two requirements. First, the drift Hamiltonian $\hat H_{d,0}$ should be diagonal (with eigenvalues denoted as $E_k$ for eigenstates $\ket{k}$) and contain certain type of nonlinearity, specifically, the nearest energy difference $\mu_k := E_k - E_{k+1}$ should satisfy $\mu_0 \neq 0$ and $\mu_k^2 \neq \mu_0^2$ for $k>0$ (or similarly $\mu_{N-2} \neq 0$ and $\mu_k^2 \neq \mu_{N-2}^2$ for $k < N-2$). Second, one of the control parts $\hat H_{c,j}$ should only have couplings between  $\ket{k}$ and $\ket{k+1}$ for all $0\leq k < N-1$. If both these conditions are satisfied, then the generated dynamical Lie group will be U($N$) when $\Tr[\hat H_d] \neq 0$ and SU($N$) otherwise. 

We can easily verify that those requirements for U($N$) group (where $N=r+1$ in our case) generation are satisfied with our $\hat H_{d,0}$ and $\hat H_{c,R}$ as long as $r \neq -\frac{2\Delta_0}{\chi} + 1$. This allows us to fix $\Delta_0 = 0$ for $r\geq 2$ in the rest of the main text.
Moreover, the two control degrees of freedom $\hat H_{c,R}$ and $\hat H_{c,I}$ provide the possibility to do any gate (up to a global phase) in an arbitrarily fast manner, as these two are also sufficient to generate full SU($N$) group~\cite{supp}. Finally, as $r$ is also adjustable, we can choose any blockade dimension we want. Therefore, unitaries defined in any finite dimension are in principle achievable.

\textit{Optimal control}--
The generalized blockade phenomena allow one, in principle, to perform an arbitrary unitary operation in an arbitrarily chosen $N$-dimensional blockaded subspace in a time much faster than $1/\chi$. The question that we now address is how to design a particular control $\alpha(t)$ to realize a target unitary. In contrast to the conventional setting, where one optimizes the control in the rotating frame of the drive~\cite{Ashhab2022}, we consider $\hat H_\dr$ defined in the instantaneous displaced rotating frame and optimize $\alpha(t)$, the frame parameter. Consequently, finding an optimal $\alpha(t)$ directly determines the corresponding physical parameters  $\Lambda_{1(2)}(t)$ and $\omega_1(t)$ via Eq.~\eqref{omega1} required for implementing the desired operations in the laboratory. In other words, by hard coding the blockade condition in the evolution through Eq.~\eqref{omega1}, we simplify the optimization task to finding $\alpha(t)$ in the small blockade Hilbert space. In this way, even though the required drive amplitudes $\Lambda_{1(2)}(t)$ can be very large, we do not need to consider large photon number states in the optimization. In this section, we only optimize the controls for a fixed evolution time $T$ in the absence of loss. We further consider the effects of loss and a realistic experimental constraint on total input power (see~\cite{supp}); as we show, these additional features lead to there being an optimal choice of gate time $T$.



Typically, optimal control algorithms such as GRAPE~\cite{Khaneja2005,Machnes2011} discretize the control pulse $\alpha(t)$, and maximize a figure of merit such as the fidelity by performing gradient-based optimization on the control parameters, i.e., the amplitude of the control at discrete time points. This standard approach would yield a piece-wise constant $\alpha(t)$, something that is highly problematic for our setup: discontinuous jumps in $\alpha(t)$ would require infinite driving power to implement, as the 1-photon driving amplitude has a term proportional to $i\dot\alpha(t)$ (c.f. Eq.~\eqref{omega1}). 

To solve this issue, in contrast to conventional methods, we expand $\alpha(t)$ on a basis of continuous functions. Since it is desirable to be in a non-displaced frame, i.e. $\alpha(0)=\alpha(T)=0$ in the beginning ($t=0$) and at the end ($t=T$) of the protocol, we use the following sine-basis ansatz for the control pulse 
\begin{equation}\label{alpha_t}
    \alpha(t) = \sum_{k=1}^{k_{\max}} \alpha_k \sin(\frac{k\pi t}{T}).
\end{equation}
Here, $k_{\max}$, the cutoff number for the highest harmonics that we use, is a hyperparameter that is chosen according to the complexity of the task, and $\alpha_k$ are complex-valued optimization variables.
In practice, we choose $k_{\max}$ heuristically in the optimization procedure. If the fidelity achieved from the optimal pulses is lower than our target, we increase $k_{\max}$ for better performance.
To implement a target unitary operation in $N$ dimensions, we maximize $F_{\rm{u}}(\{\alpha_k\})=\abs{\Tr[\hat U_\tar^\da \hat U(T)]}^2/N^2$, where $\hat{U}(T)$ is obtained by solving $\frac{\dif\ }{\dif t} \hat U(t) = -i \hat H_\dr[\alpha(t)] \hat U(t)$ for $t=T$ with the initial condition $\hat{U}(0)=\hat{I}$. When optimizing $F_{\rm{u}}$, we implicitly ignore the irrelevant global phase. Note that the dependence of the objective on $\{\alpha_k\}$ originates from the dependence of $\hat{H}_{\dr}$ on $\alpha(t)$ according to Eq.~\eqref{Hdr}. Moreover, while obtaining $F_{\rm{u}}$ involves solving an ordinary differential equation, it is still differentiable and its gradient with respect to $\{\alpha_k\}$ can be calculated using the chain rule and the adjoint sensitivity method~\cite{pontryagin1962mathematical}. Therefore, we can use gradient-based optimization to find locally optimal $\{\alpha_k\}$.

\begin{figure}[t!]
\centering
\includegraphics[scale = 0.44]{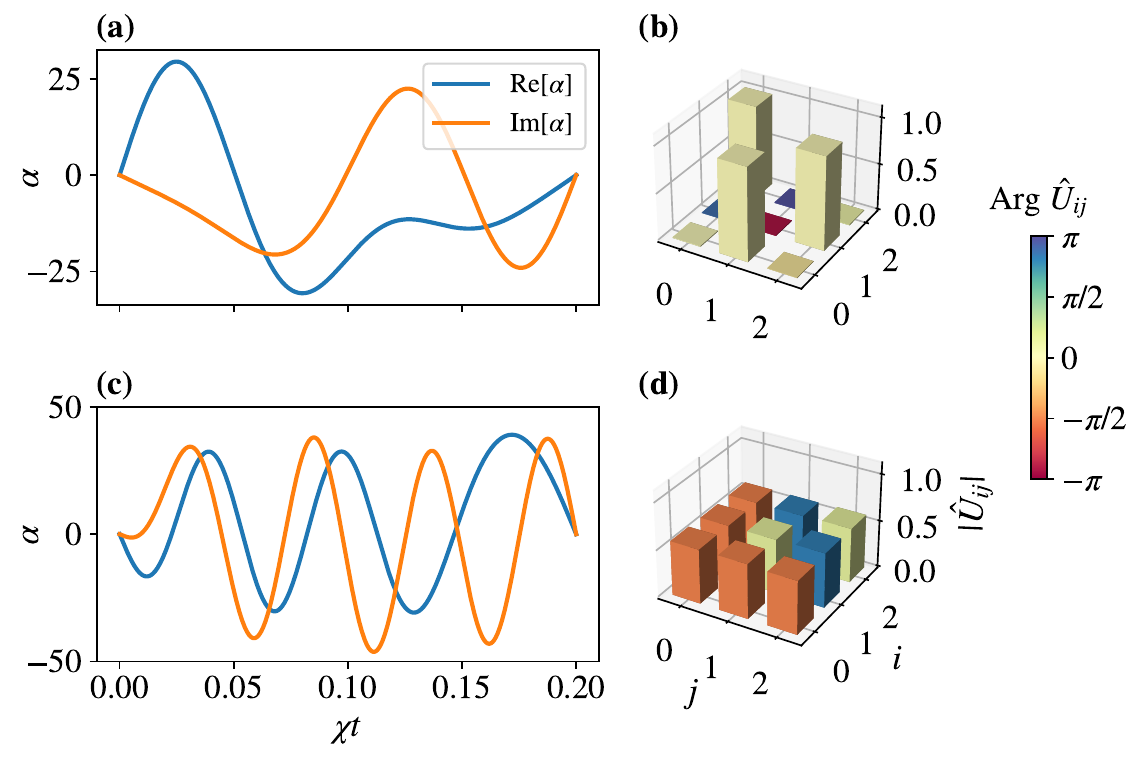}
\caption{ Optimal control for implementing arbitrary unitary operations. (a) The optimized $\alpha(t)$ function that implements permutation \eqref{eq:permu} in a 3-level blockade subspace with $k_{\max}=5$ harmonics. Total evolution time $T=0.2/\chi$. (b) The elements of the unitary $\hat U(T)$ generated by $\alpha(t)$ shown in (a) under the Fock basis, i.e., $\hat U_{ij} = \bra{i}\hat U(T)\ket{j}$.  The height and color of each bar show the absolute value and argument of the corresponding matrix elements. (c) Optimal $\alpha(t)$ for implementing Fourier transform \eqref{eq:fourtrans} in the same blockade subspace with $k_{\max}=8$ harmonics and $T=0.2/\chi$. (d) the matrix elements of $\hat U(T)$ generated by $\alpha(t)$ shown in (c).}
    \label{fig:Qutrit}
\end{figure}

To illustrate the universal controllability of our scheme, we consider the problem of implementing the permutation $\hat U_{\text{P}}$ or the Fourier transformation $\hat U_{\text{FT}}$ in a 3-level blockade subspace up to a global phase spanned by $\{\ket{m}\}_{m=0}^2$, where
\begin{subequations}
\begin{alignat}{2}
    &\hat U_{\text{P}} = \ketbra{2}{0} + \ketbra{0}{1} + \ketbra{1}{2},\label{eq:permu}\\
    &\hat U_{\text{FT}} = \frac{1}{\sqrt{3}} \sum_{m,n} e^{i\frac{2\pi m n}{3}}\ketbra{m}{n}\label{eq:fourtrans}.
\end{alignat}
\end{subequations}
We use the automatic differentiation toolbox of JAX~\cite{Jax2018}, a numerical computing package, to perform the gradient-based optimization. We choose the evolution time $T=0.2/\chi$. We also fix $\alpha(0) = \alpha(T) = 0$ and $\Delta_0 = 0$ in $\hat H_\dr$ and find the pulses $\alpha(t)$ that implements the two unitary operations of interest. In both cases, the algorithm finds a solution such that $\abs{\Tr[\hat U_\tar^\da \hat U(T)]}^2/N^2 > 1 - 10^{-4}$ (see Fig.~\ref{fig:Qutrit}).

\textit{Only 1-photon drive}--
So far, we have focused on the fundamental questions about controllability by explicitly constructing the 1- and 2-photon drives to achieve fast universal control for weak-Kerr systems in the presence of loss. Here, we move away from the ideal scenario and discuss issues relevant to experimental implementations. One question to address is whether one truly needs a distinct 2-photon drive. While this can be done in some platforms, e.g., superconducting qubits with flux-pumping~\cite{eriksson2023universal}, it can be challenging in other platforms. Here, we present a method that allows our control scheme to be implemented without any explicit independent 2-photon drive. As we show, this idea is intrinsically connected to squeezing by double-pumping a Kerr resonator~\cite{Kamal2009}.

In the absence of an independent 2-photon drive,  $\Lambda_2(t) = 0$ in Eq.~\eqref{eqn:ham_lab}. Again, we  set $\Delta_0 =0$ and choose the 1-photon driving amplitude such that in a frame rotating with $\omega_1(t)$ and displaced by $\tilde \alpha(t)$, we have the same blockade drive. In this frame, the Hamiltonian is  
\begin{equation}\label{eqn:new_single_drive}
\begin{split}
    \hat H'_\dr[\tilde{\alpha}(t)] ={}& \frac{\chi}{2} \hat a^{\da 2} \hat a^2 + [\chi \tilde\alpha(t) \hat a^\da (\hat n - r) + \hc]\\
    &+ \left[\frac{\chi}{2}\tilde\alpha^{2}(t) \hat a^{\da 2} + \hc\right],
\end{split}
\end{equation}
where the 1-photon driving amplitude is chosen to have the form
\begin{equation}\label{eq:Lambda1_tilde}
    \Lambda'_1(t) = \chi\tilde\alpha(t)[|\tilde\alpha(t)|^2 - r] + i\dot {\tilde\alpha}(t) + i\kappa\at/2.
\end{equation}
While the nonlinear single-photon drive in Eq.~\eqref{eqn:new_single_drive} has the correct form, the induced two-photon drive term in the last line violates the desired blockade condition.  

A key observation in our strategy to revive the blockade condition is that the desired fully blockaded Hamiltonian can be written as $\hat H_\dr[\alpha(t)] = \frac{1}{\sqrt{2}}\{\hat H'_\dr[e^{-\frac{i\pi}{4}}\alpha(t)] + \hat H'_\dr[e^{\frac{i\pi}{4}}\alpha(t)]\}$. Intuitively, by alternating the phase of $\tilde\alpha(t)$, we can average away and cancel the unwanted 2-photon drive term while retaining the desired nonlinear blockade drive. This observation combined with the Trotter formula 
\begin{equation}\label{Trotter}
\begin{split}
    e^{-i\frac{\hat H_1 + \hat H_2}{2} \delta T} ={}& e^{-i\hat H_1\delta T/4}e^{-i\hat H_2\delta T/2}e^{-i\hat H_1\delta T/4}\\
    & + O[(\delta T)^3]
\end{split}
\end{equation}
suggests that setting $\hat H_1(t)=\hat H'_\dr[e^{-\frac{i\pi}{4}}\alpha(t)]$ and $\hat H_2(t)=\hat H'_\dr[e^{\frac{i\pi}{4}}\alpha(t)]$ and alternating the evolution for $\delta T/2$ between the two Hamiltonians suppress the errors in violating the blockade condition to $O[(\delta T)^3]$~\cite{high_trotter}. 
However, this Trotter scheme requires discretizing $\alpha(t)$ into intervals of length $\delta T$ and implementing instantaneous displacements between $e^{\pm\frac{i\pi}{4}}\alpha$, which in practice introduces additional complexities. 


Inspired by the discrete version of the Trotter formula, we design its continuous counterpart via Magnus expansion~\cite{supp}
\begin{equation}\label{contiTrot}
\begin{split}
    \exp[-i\int_0^{\delta T} \hat H(t) \dif t] ={}& \mathcal{T}\exp[-i\int_0^{\delta T} \hat H(t) \dif t]\\
    &+ O[(\delta T)^3],
\end{split}
\end{equation}
where $\mathcal{T}$ is the time-ordering operator and $\hat{H}(t)$ for $t\in [0,\delta T]$ is chosen to be symmetric around $\delta T/2$, i.e., $\hat H(t) = \hat H(\delta T - t)$. 
As a result, we construct a new function $\tilde \alpha(t)$ that oscillates rapidly with time. 
Specifically, Eq.~\eqref{contiTrot} gives us a recipe for finding $\tilde{\alpha}(t)$ such that the coarse-grained evolution under $\hat H'_\dr[\tilde{\alpha}(t)]$ over an interval of $\delta T$ is close to that under $\hat H_\dr[\alpha(t)]$. This translates to having the average of $\tilde \alpha(t)$  over a $\delta T$ time interval centered on time $t$ to satisfy
\begin{equation}\label{eq:alphaconstraints}
    \overline{\tilde \alpha^2(t)} = 0, \quad \overline{\tilde \alpha(t)} = \alpha(t),
\end{equation}
where the overline denotes the coarse-graining time average, and $\alpha(t)$ is the optimal choice of function in $\hat H_\dr$ defined in Eq.~\eqref{Hdr} that generates the desired target unitary operation. To satisfy these constraints, we propose using the ansatz $\tilde \alpha(t) = \alpha(t) f(t)$, where $f(t)$ is a periodic function with period $\delta T = \frac{T}{M}$. Here $T$ is the total evolution time and $M$ is the number of periods during the evolution. We also denote $\omega_r := \frac{2\pi M}{T}$ for further use. Besides, to keep the same structure as the Trotter formula in Eq.~\eqref{contiTrot}, we further require that
\begin{equation}\label{f_even}
    f(t) = f(\delta T - t).
\end{equation}
Consequently, the constraints for $\tilde{\alpha}(t)$ in Eq.~\eqref{eq:alphaconstraints} translate to constraints 
\begin{equation}\label{f_avg}
    \overline{f(t)} = 1, \quad \overline{f^2(t)} = 0
\end{equation}
on $f(t)$ over over each period. This ensures that the overall evolution under $H'_\dr[\tilde \alpha(t)]$ and $\hat H_\dr[\alpha(t)]$ for time $T$ closely resemble each other, that is 
\begin{equation}\label{UniApprox}
\begin{split}
    &\mathcal{T}\exp\left\{-i \int_0^T \hat H'_\dr[\tilde \alpha(t)] \dif t\right\}\\
    \approx{}& \mathcal{T}\exp\left\{-i \int_0^T \hat H_\dr[\alpha(t)] \dif t\right\}. 
\end{split}
\end{equation}

\begin{figure}[t!]
\centering
\includegraphics[scale = 0.5]{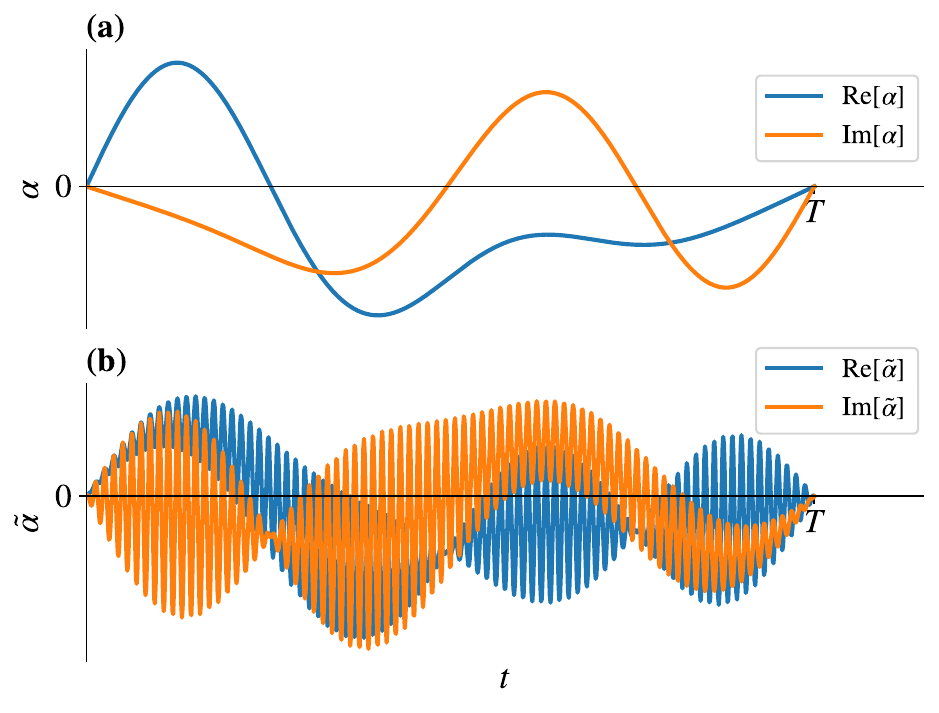}
\caption{Schematic comparison between the (a) original optimal pulse $\alpha(t)$ and (b) the modulated pulse $\at(t) = \alpha(t) f_\dbp(t)$ using Eq.~\eqref{eq:double_pump} needed in the absence of a 2-photon drive. In this example, we choose $M = 80$ periods for the modulation.}
    \label{fig:alpha_tl}
\end{figure}

One possible choice of $f(t)$ is the following
\begin{equation}\label{eq:double_pump}
    f_\dbp(t) = 1 + i\sqrt{2} \cos(\omega_r t).
\end{equation}
An example of the shape of $\at(t)$ after implementing this modulation is shown in Fig.~\ref{fig:alpha_tl} (and schematic illustration in Fig.~\ref{fig:Schemetic}(b) as well). To obtain physical intuition on the underlying mechanism, we can split $f_\dbp(t)$ into two parts. The constant part plays the role of $\Lambda_1$ in Eq.~\eqref{eqn:ham_lab}, and provides the desired nonlinear blockade drive in the displaced rotating frame.  In contrast, the time-dependent term, $i\sqrt{2} \cos(\omega_r t)$, corresponds to the double-pumping scheme with driving frequencies $\omega_1 \pm \omega_r$, as it contributes to $\Lambda'_1(t)$ via $i\dot\at(t)$. This double-pumping will effectively generate the 2-photon driving $\Lambda_2$ in Eq.~\eqref{eqn:ham_lab} from the Kerr interaction, as a result of 4-wave mixing~\cite{Kamal2009}.

The approximation in Eq.~\eqref{UniApprox} is valid if the number of periods $M$ is large enough such that (i) the error from Trotter expansion (see Eq.~(\ref{Trotter},\ref{contiTrot})) is small, and (ii) within each period of $f(t)$, the control function $\alpha(t)$ varies slowly so that the time average of $\tilde \alpha(t)$ deviates little from $\alpha(t)$.  In the infinite $M$ limit the two sides of Eq.~\eqref{UniApprox} are identical. The induced error due to the inaccuracy of Trotter approximation is analyzed in detail in Ref.~\cite{supp}.



\textit{Discussion}--
In the previous section, we have demonstrated that in principle a distinct 2-photon drive is unnecessary for universal control and explicitly provided the modified 1-photon pulse design for targeted operations. However, our operations may suffer from coherent errors that come from inaccurate Trotter approximation. This, as well as the incoherent photon loss during gate execution, serve as two sources for gate infidelity. 

If the 1-photon driving amplitude $|\Lambda_1|$ (or the input power $P_\inp$) is unlimited, both error sources can be sufficiently suppressed. We can increase the amplitude of $\alpha(t)$ so that the gate time $T$ as well as the photon loss probability is reduced. We can also increase the displaced frame oscillating frequency $\omega_r$ to reduce Trotter errors. However, if $P_\inp$ is limited, we have to make a trade-off to achieve optimized operation fidelity. In Ref.~\cite{supp}, we analyze the effect of these errors in detail for the case of preparing a single photon Fock state and show that the state preparation infidelity is given by 
\begin{equation}
    \epsilon_\tot \simeq C (\kappa_e + \kappa_i) T + \frac{C'}{P_{\rm{in}}^2 \kappa_e^2 \chi^{10} T^{14}},
\end{equation}
where $\kappa_{e(i)}$ stands for the external (internal) loss rate. Here the first term captures the effect of loss errors growing with time $T$, and the second term captures the Trotter errors that scale inverse polynomially with $T$. Therefore, there exist an optimal time and external loss rate for a given power and internal loss rate that minimizes the total error, which scales as $\epsilon_\tot^\opt \propto \kappa_i^{4/5}/(P_\inp^{2/15}\chi^{2/3})$. 
For example, suppose we have a niobium nitride resonator~\cite{Anferov2020} with $\omega_c = 2\pi \times 100$ GHz, $\chi = -1$ kHz and an improved $|\chi|/\kappa_i = 0.5$ in the future. Then, to prepare $\ket{1}$ state with 90\% fidelity the required $P_\inp$ is around 30 nW. The corresponding required $\alpha\approx 15$. These numbers suggest that our scheme may work well in the regime that self-Kerr is comparable to or slightly lower than the loss rate. However, if $\kappa_i$ is increased by a factor of 10 we need an increase of $P_\inp$ by a factor of $10^{6}$ to keep the same $\epsilon_\tot^\opt$, which indicates that our protocol may not be power-friendly in the large $\kappa_i$ regime. On the other hand, if we can implement the 2-photon drive directly which is also power-efficient so that we only need to consider the power cost for the 1-photon drive, then the error scaling will be modified as $\epsilon\propto \kappa_i^{5/6}/(P_\inp^{1/6}\chi^{2/3})$. We leave the investigation of other power-saving protocols for universal control as further work.

We also investigate the optimality of the Trotter scheme by comparing it with a direct optimization scheme.  Specifically, we assume that the two-photon drive is absent from the beginning, and directly maximize the fidelity by optimizing $\alpha(t)$ using a more expressive ansatz that utilizes neural networks with periodic activation functions~\cite{sitzmann2020implicit}. Our results indicate that this heuristic approach does not yield a better solution (in terms of fidelity and required power) compared to the Trotter scheme. We refer the reader to \cite{supp} for more details.

Additionally, if our 1-photon drive is too strong, the RWA may not be valid since the frequency $\omega_c$ of the resonator is always finite. The counter-rotating (non-RWA) terms like $(\Lambda_1^*(t) e^{i\omega_1(t)t}\hat a^\da + \hc)$ should be considered in Eq.~\eqref{eqn:ham_lab}. In Ref.~\cite{supp}, we discuss the way to mitigate the dominant effects from non-RWA terms by adjusting the driving amplitude $\Lambda_1(t)$ and frequency $\omega_1(t)$. It is also worth mentioning that there will be no $(\Lambda_1^*(t) e^{i\omega_1(t)t}\hat a^\da + \hc)$ term if on the hardware we can drive both charge and flux simultaneously with properly chosen amplitudes, which provides another way to mitigate non-RWA effects. However, there could still be a fundamental limitation on the fidelity as a function of two dimensionless parameters $\kappa_i/\chi$ and internal quality factor $Q_i := \omega_c/\kappa_i$. For the task of preparing $\ket{1}$, we found a rough lower bound for infidelity as $\epsilon \gtrsim 3\pi (\kappa_i/\chi)^{2/3}/(16Q_i^{1/3})$ even if we can directly implement 2-photon drive or do 1-photon drives on both charge and flux quadratures~\cite{supp}. We leave the further improvement of this bound as an open question.

\textit{Conclusion}--
We show that by making a non-trivial extension of the  displacement-induced, weak-Kerr photon blockade
of Ref.~\cite{Lingenfelter2021}, one can achieve any unitary operation in a  blockade subspace of arbitrary dimension. The speed of the operations can be enhanced by using large displacements to overcome the adverse effects of photon loss.  Further, this can be implemented using only 1-photon drives, provided that the input power is not a limitation and RWA conditions are justified.
Moreover, from a computational perspective, our work simplifies the task of optimal control in such systems. It reduces the computational overhead for finding control sequences that utilize large photon number states by working in a special instantaneously displaced frame. As a result, our work provides a novel and efficient quantum control protocol for weak nonlinear bosonic systems, which could be helpful for future quantum information processing tasks on suitable platforms.


\begin{acknowledgments}

We thank Alexander Anferov, Srivatsan Chakram, Dirk Englund, Kevin He, Mikkel Heuck, Kurt Jacobs, Stefan Krastanov, Kan-Heng Lee, Hong Tang, Zhaoyou Wang, and Mingrui Xu for helpful discussion. We acknowledge support from the ARO (W911NF-23-1-0077), ARO MURI (W911NF-21-1-0325), AFOSR MURI (FA9550-19-1-0399, FA9550-21-1-0209, FA9550-23-1-0338), NSF (OMA-1936118, ERC-1941583, OMA-2137642, OSI-2326767, CCF-2312755), NTT Research, the Simons Foundation (Grant No. 669487), and the Packard Foundation (2020-71479). This material is based upon work supported by the U.S. Department of Energy, Office of Science, and National Quantum Information Science Research Centers.
\end{acknowledgments}


\nocite{*}

\bibliography{ref}

\begin{thebibliography}{32}%
\makeatletter
\providecommand \@ifxundefined [1]{%
 \@ifx{#1\undefined}
}%
\providecommand \@ifnum [1]{%
 \ifnum #1\expandafter \@firstoftwo
 \else \expandafter \@secondoftwo
 \fi
}%
\providecommand \@ifx [1]{%
 \ifx #1\expandafter \@firstoftwo
 \else \expandafter \@secondoftwo
 \fi
}%
\providecommand \natexlab [1]{#1}%
\providecommand \enquote  [1]{``#1''}%
\providecommand \bibnamefont  [1]{#1}%
\providecommand \bibfnamefont [1]{#1}%
\providecommand \citenamefont [1]{#1}%
\providecommand \href@noop [0]{\@secondoftwo}%
\providecommand \href [0]{\begingroup \@sanitize@url \@href}%
\providecommand \@href[1]{\@@startlink{#1}\@@href}%
\providecommand \@@href[1]{\endgroup#1\@@endlink}%
\providecommand \@sanitize@url [0]{\catcode `\\12\catcode `\$12\catcode
  `\&12\catcode `\#12\catcode `\^12\catcode `\_12\catcode `\%12\relax}%
\providecommand \@@startlink[1]{}%
\providecommand \@@endlink[0]{}%
\providecommand \url  [0]{\begingroup\@sanitize@url \@url }%
\providecommand \@url [1]{\endgroup\@href {#1}{\urlprefix }}%
\providecommand \urlprefix  [0]{URL }%
\providecommand \Eprint [0]{\href }%
\providecommand \doibase [0]{http://dx.doi.org/}%
\providecommand \selectlanguage [0]{\@gobble}%
\providecommand \bibinfo  [0]{\@secondoftwo}%
\providecommand \bibfield  [0]{\@secondoftwo}%
\providecommand \translation [1]{[#1]}%
\providecommand \BibitemOpen [0]{}%
\providecommand \bibitemStop [0]{}%
\providecommand \bibitemNoStop [0]{.\EOS\space}%
\providecommand \EOS [0]{\spacefactor3000\relax}%
\providecommand \BibitemShut  [1]{\csname bibitem#1\endcsname}%
\let\auto@bib@innerbib\@empty
\bibitem [{\citenamefont {Joshi}\ \emph {et~al.}(2021)\citenamefont {Joshi},
  \citenamefont {Noh},\ and\ \citenamefont {Gao}}]{joshi2021quantum}%
  \BibitemOpen
  \bibfield  {author} {\bibinfo {author} {\bibfnamefont {A.}~\bibnamefont
  {Joshi}}, \bibinfo {author} {\bibfnamefont {K.}~\bibnamefont {Noh}}, \ and\
  \bibinfo {author} {\bibfnamefont {Y.~Y.}\ \bibnamefont {Gao}},\ }\href
  {\doibase 10.1088/2058-9565/abe989} {\bibfield  {journal} {\bibinfo
  {journal} {Quantum Science and Technology}\ }\textbf {\bibinfo {volume}
  {6}},\ \bibinfo {pages} {033001} (\bibinfo {year} {2021})}\BibitemShut
  {NoStop}%
\bibitem [{\citenamefont {Kok}\ \emph {et~al.}(2007)\citenamefont {Kok},
  \citenamefont {Munro}, \citenamefont {Nemoto}, \citenamefont {Ralph},
  \citenamefont {Dowling},\ and\ \citenamefont {Milburn}}]{kok_linear_2007}%
  \BibitemOpen
  \bibfield  {author} {\bibinfo {author} {\bibfnamefont {P.}~\bibnamefont
  {Kok}}, \bibinfo {author} {\bibfnamefont {W.~J.}\ \bibnamefont {Munro}},
  \bibinfo {author} {\bibfnamefont {K.}~\bibnamefont {Nemoto}}, \bibinfo
  {author} {\bibfnamefont {T.~C.}\ \bibnamefont {Ralph}}, \bibinfo {author}
  {\bibfnamefont {J.~P.}\ \bibnamefont {Dowling}}, \ and\ \bibinfo {author}
  {\bibfnamefont {G.~J.}\ \bibnamefont {Milburn}},\ }\href {\doibase
  10.1103/RevModPhys.79.135} {\bibfield  {journal} {\bibinfo  {journal}
  {Reviews of Modern Physics}\ }\textbf {\bibinfo {volume} {79}},\ \bibinfo
  {pages} {135} (\bibinfo {year} {2007})}\BibitemShut {NoStop}%
\bibitem [{\citenamefont {Lloyd}\ and\ \citenamefont
  {Braunstein}(1999)}]{Lloyd1999}%
  \BibitemOpen
  \bibfield  {author} {\bibinfo {author} {\bibfnamefont {S.}~\bibnamefont
  {Lloyd}}\ and\ \bibinfo {author} {\bibfnamefont {S.~L.}\ \bibnamefont
  {Braunstein}},\ }\href {\doibase 10.1103/physrevlett.82.1784} {\bibfield
  {journal} {\bibinfo  {journal} {Physical Review Letters}\ }\textbf {\bibinfo
  {volume} {82}},\ \bibinfo {pages} {1784} (\bibinfo {year}
  {1999})}\BibitemShut {NoStop}%
\bibitem [{\citenamefont {Eriksson}\ \emph {et~al.}()\citenamefont {Eriksson},
  \citenamefont {Sépulcre}, \citenamefont {Kervinen}, \citenamefont
  {Hillmann}, \citenamefont {Kudra}, \citenamefont {Dupouy}, \citenamefont
  {Lu}, \citenamefont {Khanahmadi}, \citenamefont {Yang}, \citenamefont
  {Moreno}, \citenamefont {Delsing},\ and\ \citenamefont
  {Gasparinetti}}]{eriksson2023universal}%
  \BibitemOpen
  \bibfield  {author} {\bibinfo {author} {\bibfnamefont {A.~M.}\ \bibnamefont
  {Eriksson}}, \bibinfo {author} {\bibfnamefont {T.}~\bibnamefont {Sépulcre}},
  \bibinfo {author} {\bibfnamefont {M.}~\bibnamefont {Kervinen}}, \bibinfo
  {author} {\bibfnamefont {T.}~\bibnamefont {Hillmann}}, \bibinfo {author}
  {\bibfnamefont {M.}~\bibnamefont {Kudra}}, \bibinfo {author} {\bibfnamefont
  {S.}~\bibnamefont {Dupouy}}, \bibinfo {author} {\bibfnamefont
  {Y.}~\bibnamefont {Lu}}, \bibinfo {author} {\bibfnamefont {M.}~\bibnamefont
  {Khanahmadi}}, \bibinfo {author} {\bibfnamefont {J.}~\bibnamefont {Yang}},
  \bibinfo {author} {\bibfnamefont {C.~C.}\ \bibnamefont {Moreno}}, \bibinfo
  {author} {\bibfnamefont {P.}~\bibnamefont {Delsing}}, \ and\ \bibinfo
  {author} {\bibfnamefont {S.}~\bibnamefont {Gasparinetti}},\ }\href@noop {}
  {}\Eprint {http://arxiv.org/abs/2308.15320} {arXiv:2308.15320 [quant-ph]}
  \BibitemShut {NoStop}%
\bibitem [{\citenamefont {Krastanov}\ \emph {et~al.}(2015)\citenamefont
  {Krastanov}, \citenamefont {Albert}, \citenamefont {Shen}, \citenamefont
  {Zou}, \citenamefont {Heeres}, \citenamefont {Vlastakis}, \citenamefont
  {Schoelkopf},\ and\ \citenamefont {Jiang}}]{Krastanov2015}%
  \BibitemOpen
  \bibfield  {author} {\bibinfo {author} {\bibfnamefont {S.}~\bibnamefont
  {Krastanov}}, \bibinfo {author} {\bibfnamefont {V.~V.}\ \bibnamefont
  {Albert}}, \bibinfo {author} {\bibfnamefont {C.}~\bibnamefont {Shen}},
  \bibinfo {author} {\bibfnamefont {C.-L.}\ \bibnamefont {Zou}}, \bibinfo
  {author} {\bibfnamefont {R.~W.}\ \bibnamefont {Heeres}}, \bibinfo {author}
  {\bibfnamefont {B.}~\bibnamefont {Vlastakis}}, \bibinfo {author}
  {\bibfnamefont {R.~J.}\ \bibnamefont {Schoelkopf}}, \ and\ \bibinfo {author}
  {\bibfnamefont {L.}~\bibnamefont {Jiang}},\ }\href {\doibase
  10.1103/physreva.92.040303} {\bibfield  {journal} {\bibinfo  {journal}
  {Physical Review A}\ }\textbf {\bibinfo {volume} {92}},\ \bibinfo {pages}
  {040303(R)} (\bibinfo {year} {2015})}\BibitemShut {NoStop}%
\bibitem [{\citenamefont {Hacohen-Gourgy}\ \emph {et~al.}(2016)\citenamefont
  {Hacohen-Gourgy}, \citenamefont {Martin}, \citenamefont {Flurin},
  \citenamefont {Ramasesh}, \citenamefont {Whaley},\ and\ \citenamefont
  {Siddiqi}}]{Hacohen-Gourgy2016}%
  \BibitemOpen
  \bibfield  {author} {\bibinfo {author} {\bibfnamefont {S.}~\bibnamefont
  {Hacohen-Gourgy}}, \bibinfo {author} {\bibfnamefont {L.~S.}\ \bibnamefont
  {Martin}}, \bibinfo {author} {\bibfnamefont {E.}~\bibnamefont {Flurin}},
  \bibinfo {author} {\bibfnamefont {V.~V.}\ \bibnamefont {Ramasesh}}, \bibinfo
  {author} {\bibfnamefont {K.~B.}\ \bibnamefont {Whaley}}, \ and\ \bibinfo
  {author} {\bibfnamefont {I.}~\bibnamefont {Siddiqi}},\ }\href {\doibase
  10.1038/nature19762} {\bibfield  {journal} {\bibinfo  {journal} {Nature}\
  }\textbf {\bibinfo {volume} {538}},\ \bibinfo {pages} {491–494} (\bibinfo
  {year} {2016})}\BibitemShut {NoStop}%
\bibitem [{\citenamefont {Eickbusch}\ \emph {et~al.}(2022)\citenamefont
  {Eickbusch}, \citenamefont {Sivak}, \citenamefont {Ding}, \citenamefont
  {Elder}, \citenamefont {Jha}, \citenamefont {Venkatraman}, \citenamefont
  {Royer}, \citenamefont {Girvin}, \citenamefont {Schoelkopf},\ and\
  \citenamefont {Devoret}}]{Eickbusch2022}%
  \BibitemOpen
  \bibfield  {author} {\bibinfo {author} {\bibfnamefont {A.}~\bibnamefont
  {Eickbusch}}, \bibinfo {author} {\bibfnamefont {V.}~\bibnamefont {Sivak}},
  \bibinfo {author} {\bibfnamefont {A.~Z.}\ \bibnamefont {Ding}}, \bibinfo
  {author} {\bibfnamefont {S.~S.}\ \bibnamefont {Elder}}, \bibinfo {author}
  {\bibfnamefont {S.~R.}\ \bibnamefont {Jha}}, \bibinfo {author} {\bibfnamefont
  {J.}~\bibnamefont {Venkatraman}}, \bibinfo {author} {\bibfnamefont
  {B.}~\bibnamefont {Royer}}, \bibinfo {author} {\bibfnamefont {S.~M.}\
  \bibnamefont {Girvin}}, \bibinfo {author} {\bibfnamefont {R.~J.}\
  \bibnamefont {Schoelkopf}}, \ and\ \bibinfo {author} {\bibfnamefont {M.~H.}\
  \bibnamefont {Devoret}},\ }\href {\doibase 10.1038/s41567-022-01776-9}
  {\bibfield  {journal} {\bibinfo  {journal} {Nature Physics}\ }\textbf
  {\bibinfo {volume} {18}},\ \bibinfo {pages} {1464–1469} (\bibinfo {year}
  {2022})}\BibitemShut {NoStop}%
\bibitem [{\citenamefont {Diringer}\ \emph {et~al.}()\citenamefont {Diringer},
  \citenamefont {Blumenthal}, \citenamefont {Grinberg}, \citenamefont {Jiang},\
  and\ \citenamefont {Hacohen-Gourgy}}]{diringer_conditional_2023}%
  \BibitemOpen
  \bibfield  {author} {\bibinfo {author} {\bibfnamefont {A.~A.}\ \bibnamefont
  {Diringer}}, \bibinfo {author} {\bibfnamefont {E.}~\bibnamefont
  {Blumenthal}}, \bibinfo {author} {\bibfnamefont {A.}~\bibnamefont
  {Grinberg}}, \bibinfo {author} {\bibfnamefont {L.}~\bibnamefont {Jiang}}, \
  and\ \bibinfo {author} {\bibfnamefont {S.}~\bibnamefont {Hacohen-Gourgy}},\
  }\href@noop {} {}\Eprint {http://arxiv.org/abs/2301.09831} {arXiv:2301.09831
  [quant-ph]} \BibitemShut {NoStop}%
\bibitem [{\citenamefont {Chan}\ \emph {et~al.}(2011)\citenamefont {Chan},
  \citenamefont {Alegre}, \citenamefont {Safavi-Naeini}, \citenamefont {Hill},
  \citenamefont {Krause}, \citenamefont {Gröblacher}, \citenamefont
  {Aspelmeyer},\ and\ \citenamefont {Painter}}]{chan_laser_2011}%
  \BibitemOpen
  \bibfield  {author} {\bibinfo {author} {\bibfnamefont {J.}~\bibnamefont
  {Chan}}, \bibinfo {author} {\bibfnamefont {T.~P.~M.}\ \bibnamefont {Alegre}},
  \bibinfo {author} {\bibfnamefont {A.~H.}\ \bibnamefont {Safavi-Naeini}},
  \bibinfo {author} {\bibfnamefont {J.~T.}\ \bibnamefont {Hill}}, \bibinfo
  {author} {\bibfnamefont {A.}~\bibnamefont {Krause}}, \bibinfo {author}
  {\bibfnamefont {S.}~\bibnamefont {Gröblacher}}, \bibinfo {author}
  {\bibfnamefont {M.}~\bibnamefont {Aspelmeyer}}, \ and\ \bibinfo {author}
  {\bibfnamefont {O.}~\bibnamefont {Painter}},\ }\href {\doibase
  10.1038/nature10461} {\bibfield  {journal} {\bibinfo  {journal} {Nature}\
  }\textbf {\bibinfo {volume} {478}},\ \bibinfo {pages} {89} (\bibinfo {year}
  {2011})}\BibitemShut {NoStop}%
\bibitem [{\citenamefont {Patel}\ \emph {et~al.}(2021)\citenamefont {Patel},
  \citenamefont {McKenna}, \citenamefont {Wang}, \citenamefont {Witmer},
  \citenamefont {Jiang}, \citenamefont {Van~Laer}, \citenamefont {Sarabalis},\
  and\ \citenamefont {Safavi-Naeini}}]{patel_room-temperature_2021}%
  \BibitemOpen
  \bibfield  {author} {\bibinfo {author} {\bibfnamefont {R.~N.}\ \bibnamefont
  {Patel}}, \bibinfo {author} {\bibfnamefont {T.~P.}\ \bibnamefont {McKenna}},
  \bibinfo {author} {\bibfnamefont {Z.}~\bibnamefont {Wang}}, \bibinfo {author}
  {\bibfnamefont {J.~D.}\ \bibnamefont {Witmer}}, \bibinfo {author}
  {\bibfnamefont {W.}~\bibnamefont {Jiang}}, \bibinfo {author} {\bibfnamefont
  {R.}~\bibnamefont {Van~Laer}}, \bibinfo {author} {\bibfnamefont {C.~J.}\
  \bibnamefont {Sarabalis}}, \ and\ \bibinfo {author} {\bibfnamefont {A.~H.}\
  \bibnamefont {Safavi-Naeini}},\ }\href {\doibase
  10.1103/PhysRevLett.127.133602} {\bibfield  {journal} {\bibinfo  {journal}
  {Physical Review Letters}\ }\textbf {\bibinfo {volume} {127}},\ \bibinfo
  {pages} {133602} (\bibinfo {year} {2021})}\BibitemShut {NoStop}%
\bibitem [{\citenamefont {Vernon}\ and\ \citenamefont
  {Sipe}(2015)}]{Vernon2015}%
  \BibitemOpen
  \bibfield  {author} {\bibinfo {author} {\bibfnamefont {Z.}~\bibnamefont
  {Vernon}}\ and\ \bibinfo {author} {\bibfnamefont {J.~E.}\ \bibnamefont
  {Sipe}},\ }\href {\doibase 10.1103/physreva.91.053802} {\bibfield  {journal}
  {\bibinfo  {journal} {Physical Review A}\ }\textbf {\bibinfo {volume} {91}},\
  \bibinfo {pages} {053802} (\bibinfo {year} {2015})}\BibitemShut {NoStop}%
\bibitem [{\citenamefont {Choi}\ \emph {et~al.}(2017)\citenamefont {Choi},
  \citenamefont {Heuck},\ and\ \citenamefont {Englund}}]{Choi2017}%
  \BibitemOpen
  \bibfield  {author} {\bibinfo {author} {\bibfnamefont {H.}~\bibnamefont
  {Choi}}, \bibinfo {author} {\bibfnamefont {M.}~\bibnamefont {Heuck}}, \ and\
  \bibinfo {author} {\bibfnamefont {D.}~\bibnamefont {Englund}},\ }\href
  {\doibase 10.1103/PhysRevLett.118.223605} {\bibfield  {journal} {\bibinfo
  {journal} {Physical Review Letters}\ }\textbf {\bibinfo {volume} {118}},\
  \bibinfo {pages} {223605} (\bibinfo {year} {2017})}\BibitemShut {NoStop}%
\bibitem [{\citenamefont {Anferov}\ \emph {et~al.}(2020)\citenamefont
  {Anferov}, \citenamefont {Suleymanzade}, \citenamefont {Oriani},
  \citenamefont {Simon},\ and\ \citenamefont {Schuster}}]{Anferov2020}%
  \BibitemOpen
  \bibfield  {author} {\bibinfo {author} {\bibfnamefont {A.}~\bibnamefont
  {Anferov}}, \bibinfo {author} {\bibfnamefont {A.}~\bibnamefont
  {Suleymanzade}}, \bibinfo {author} {\bibfnamefont {A.}~\bibnamefont
  {Oriani}}, \bibinfo {author} {\bibfnamefont {J.}~\bibnamefont {Simon}}, \
  and\ \bibinfo {author} {\bibfnamefont {D.~I.}\ \bibnamefont {Schuster}},\
  }\href {\doibase 10.1103/physrevapplied.13.024056} {\bibfield  {journal}
  {\bibinfo  {journal} {Physical Review Applied}\ }\textbf {\bibinfo {volume}
  {13}},\ \bibinfo {pages} {024056} (\bibinfo {year} {2020})}\BibitemShut
  {NoStop}%
\bibitem [{\citenamefont {Xu}\ \emph {et~al.}(2023)\citenamefont {Xu},
  \citenamefont {Cheng}, \citenamefont {Wu}, \citenamefont {Liu},\ and\
  \citenamefont {Tang}}]{Xu2023}%
  \BibitemOpen
  \bibfield  {author} {\bibinfo {author} {\bibfnamefont {M.}~\bibnamefont
  {Xu}}, \bibinfo {author} {\bibfnamefont {R.}~\bibnamefont {Cheng}}, \bibinfo
  {author} {\bibfnamefont {Y.}~\bibnamefont {Wu}}, \bibinfo {author}
  {\bibfnamefont {G.}~\bibnamefont {Liu}}, \ and\ \bibinfo {author}
  {\bibfnamefont {H.~X.}\ \bibnamefont {Tang}},\ }\href {\doibase
  10.1103/PRXQuantum.4.010322} {\bibfield  {journal} {\bibinfo  {journal} {PRX
  Quantum}\ }\textbf {\bibinfo {volume} {4}},\ \bibinfo {pages} {010322}
  (\bibinfo {year} {2023})}\BibitemShut {NoStop}%
\bibitem [{\citenamefont {Lingenfelter}\ \emph {et~al.}(2021)\citenamefont
  {Lingenfelter}, \citenamefont {Roberts},\ and\ \citenamefont
  {Clerk}}]{Lingenfelter2021}%
  \BibitemOpen
  \bibfield  {author} {\bibinfo {author} {\bibfnamefont {A.}~\bibnamefont
  {Lingenfelter}}, \bibinfo {author} {\bibfnamefont {D.}~\bibnamefont
  {Roberts}}, \ and\ \bibinfo {author} {\bibfnamefont {A.~A.}\ \bibnamefont
  {Clerk}},\ }\href {\doibase 10.1126/sciadv.abj1916} {\bibfield  {journal}
  {\bibinfo  {journal} {Science Advances}\ }\textbf {\bibinfo {volume} {7}},\
  \bibinfo {pages} {eabj1916} (\bibinfo {year} {2021})}\BibitemShut {NoStop}%
\bibitem [{\citenamefont {Bruch}\ \emph {et~al.}(2019)\citenamefont {Bruch},
  \citenamefont {Liu}, \citenamefont {Surya}, \citenamefont {Zou},\ and\
  \citenamefont {Tang}}]{Bruch2019}%
  \BibitemOpen
  \bibfield  {author} {\bibinfo {author} {\bibfnamefont {A.~W.}\ \bibnamefont
  {Bruch}}, \bibinfo {author} {\bibfnamefont {X.}~\bibnamefont {Liu}}, \bibinfo
  {author} {\bibfnamefont {J.~B.}\ \bibnamefont {Surya}}, \bibinfo {author}
  {\bibfnamefont {C.-L.}\ \bibnamefont {Zou}}, \ and\ \bibinfo {author}
  {\bibfnamefont {H.~X.}\ \bibnamefont {Tang}},\ }\href {\doibase
  10.1364/OPTICA.6.001361} {\bibfield  {journal} {\bibinfo  {journal} {Optica}\
  }\textbf {\bibinfo {volume} {6}},\ \bibinfo {pages} {1361} (\bibinfo {year}
  {2019})}\BibitemShut {NoStop}%
\bibitem [{\citenamefont {Kamal}\ \emph {et~al.}(2009)\citenamefont {Kamal},
  \citenamefont {Marblestone},\ and\ \citenamefont {Devoret}}]{Kamal2009}%
  \BibitemOpen
  \bibfield  {author} {\bibinfo {author} {\bibfnamefont {A.}~\bibnamefont
  {Kamal}}, \bibinfo {author} {\bibfnamefont {A.}~\bibnamefont {Marblestone}},
  \ and\ \bibinfo {author} {\bibfnamefont {M.}~\bibnamefont {Devoret}},\ }\href
  {\doibase 10.1103/PhysRevB.79.184301} {\bibfield  {journal} {\bibinfo
  {journal} {Physical Review B}\ }\textbf {\bibinfo {volume} {79}},\ \bibinfo
  {pages} {184301} (\bibinfo {year} {2009})}\BibitemShut {NoStop}%
\bibitem [{sup()}]{supp}%
  \BibitemOpen
  \href@noop {} {}\bibinfo {note} {See Supplemental Material for the detailed
  discussion on the universality proof, error analysis for our scheme, the
  estimation of the feasibility on several experimental platforms to justify
  the scheme, as well as other numerical methods to perform optimal control for
  our problem.}\BibitemShut {Stop}%
\bibitem [{\citenamefont {Boscain}\ \emph {et~al.}(2021)\citenamefont
  {Boscain}, \citenamefont {Sigalotti},\ and\ \citenamefont
  {Sugny}}]{boscain2021intro}%
  \BibitemOpen
  \bibfield  {author} {\bibinfo {author} {\bibfnamefont {U.}~\bibnamefont
  {Boscain}}, \bibinfo {author} {\bibfnamefont {M.}~\bibnamefont {Sigalotti}},
  \ and\ \bibinfo {author} {\bibfnamefont {D.}~\bibnamefont {Sugny}},\ }\href
  {\doibase 10.1103/PRXQuantum.2.030203} {\bibfield  {journal} {\bibinfo
  {journal} {PRX Quantum}\ }\textbf {\bibinfo {volume} {2}},\ \bibinfo {pages}
  {030203} (\bibinfo {year} {2021})}\BibitemShut {NoStop}%
\bibitem [{\citenamefont {Schirmer}\ \emph {et~al.}(2001)\citenamefont
  {Schirmer}, \citenamefont {Fu},\ and\ \citenamefont
  {Solomon}}]{Schirmer2001}%
  \BibitemOpen
  \bibfield  {author} {\bibinfo {author} {\bibfnamefont {S.~G.}\ \bibnamefont
  {Schirmer}}, \bibinfo {author} {\bibfnamefont {H.}~\bibnamefont {Fu}}, \ and\
  \bibinfo {author} {\bibfnamefont {A.~I.}\ \bibnamefont {Solomon}},\ }\href
  {\doibase 10.1103/physreva.63.063410} {\bibfield  {journal} {\bibinfo
  {journal} {Physical Review A}\ }\textbf {\bibinfo {volume} {63}},\ \bibinfo
  {pages} {063410} (\bibinfo {year} {2001})}\BibitemShut {NoStop}%
\bibitem [{\citenamefont {Ashhab}\ \emph {et~al.}(2022)\citenamefont {Ashhab},
  \citenamefont {Yoshihara}, \citenamefont {Fuse}, \citenamefont {Yamamoto},
  \citenamefont {Lupascu},\ and\ \citenamefont {Semba}}]{Ashhab2022}%
  \BibitemOpen
  \bibfield  {author} {\bibinfo {author} {\bibfnamefont {S.}~\bibnamefont
  {Ashhab}}, \bibinfo {author} {\bibfnamefont {F.}~\bibnamefont {Yoshihara}},
  \bibinfo {author} {\bibfnamefont {T.}~\bibnamefont {Fuse}}, \bibinfo {author}
  {\bibfnamefont {N.}~\bibnamefont {Yamamoto}}, \bibinfo {author}
  {\bibfnamefont {A.}~\bibnamefont {Lupascu}}, \ and\ \bibinfo {author}
  {\bibfnamefont {K.}~\bibnamefont {Semba}},\ }\href {\doibase
  10.1103/PhysRevA.105.042614} {\bibfield  {journal} {\bibinfo  {journal}
  {Physical Review A}\ }\textbf {\bibinfo {volume} {105}},\ \bibinfo {pages}
  {042614} (\bibinfo {year} {2022})}\BibitemShut {NoStop}%
\bibitem [{\citenamefont {Khaneja}\ \emph {et~al.}(2005)\citenamefont
  {Khaneja}, \citenamefont {Reiss}, \citenamefont {Kehlet}, \citenamefont
  {Schulte-Herbrüggen},\ and\ \citenamefont {Glaser}}]{Khaneja2005}%
  \BibitemOpen
  \bibfield  {author} {\bibinfo {author} {\bibfnamefont {N.}~\bibnamefont
  {Khaneja}}, \bibinfo {author} {\bibfnamefont {T.}~\bibnamefont {Reiss}},
  \bibinfo {author} {\bibfnamefont {C.}~\bibnamefont {Kehlet}}, \bibinfo
  {author} {\bibfnamefont {T.}~\bibnamefont {Schulte-Herbrüggen}}, \ and\
  \bibinfo {author} {\bibfnamefont {S.~J.}\ \bibnamefont {Glaser}},\ }\href
  {\doibase https://doi.org/10.1016/j.jmr.2004.11.004} {\bibfield  {journal}
  {\bibinfo  {journal} {Journal of Magnetic Resonance}\ }\textbf {\bibinfo
  {volume} {172}},\ \bibinfo {pages} {296} (\bibinfo {year}
  {2005})}\BibitemShut {NoStop}%
\bibitem [{\citenamefont {Machnes}\ \emph {et~al.}(2011)\citenamefont
  {Machnes}, \citenamefont {Sander}, \citenamefont {Glaser}, \citenamefont
  {de~Fouquières}, \citenamefont {Gruslys}, \citenamefont {Schirmer},\ and\
  \citenamefont {Schulte-Herbrüggen}}]{Machnes2011}%
  \BibitemOpen
  \bibfield  {author} {\bibinfo {author} {\bibfnamefont {S.}~\bibnamefont
  {Machnes}}, \bibinfo {author} {\bibfnamefont {U.}~\bibnamefont {Sander}},
  \bibinfo {author} {\bibfnamefont {S.~J.}\ \bibnamefont {Glaser}}, \bibinfo
  {author} {\bibfnamefont {P.}~\bibnamefont {de~Fouquières}}, \bibinfo
  {author} {\bibfnamefont {A.}~\bibnamefont {Gruslys}}, \bibinfo {author}
  {\bibfnamefont {S.}~\bibnamefont {Schirmer}}, \ and\ \bibinfo {author}
  {\bibfnamefont {T.}~\bibnamefont {Schulte-Herbrüggen}},\ }\href {\doibase
  10.1103/physreva.84.022305} {\bibfield  {journal} {\bibinfo  {journal}
  {Physical Review A}\ }\textbf {\bibinfo {volume} {84}},\ \bibinfo {pages}
  {022305} (\bibinfo {year} {2011})}\BibitemShut {NoStop}%
\bibitem [{\citenamefont {Pontryagin}\ \emph {et~al.}(1962)\citenamefont
  {Pontryagin}, \citenamefont {Bolt{\^a}nskij}, \citenamefont {Collection},
  \citenamefont {Trirogoff}, \citenamefont {Neustadt}, \citenamefont
  {Gamkrelidze},\ and\ \citenamefont
  {Mi{\^s}enko}}]{pontryagin1962mathematical}%
  \BibitemOpen
  \bibfield  {author} {\bibinfo {author} {\bibfnamefont {L.}~\bibnamefont
  {Pontryagin}}, \bibinfo {author} {\bibfnamefont {V.}~\bibnamefont
  {Bolt{\^a}nskij}}, \bibinfo {author} {\bibfnamefont {K.~M.~R.}\ \bibnamefont
  {Collection}}, \bibinfo {author} {\bibfnamefont {K.}~\bibnamefont
  {Trirogoff}}, \bibinfo {author} {\bibfnamefont {L.}~\bibnamefont {Neustadt}},
  \bibinfo {author} {\bibfnamefont {R.}~\bibnamefont {Gamkrelidze}}, \ and\
  \bibinfo {author} {\bibfnamefont {E.}~\bibnamefont {Mi{\^s}enko}},\ }\href
  {https://books.google.com/books?id=ntNSAAAAMAAJ} {\emph {\bibinfo {title}
  {The Mathematical Theory of Optimal Processes}}}\ (\bibinfo  {publisher}
  {Interscience Publishers},\ \bibinfo {year} {1962})\BibitemShut {NoStop}%
\bibitem [{\citenamefont {Bradbury}\ \emph {et~al.}(2018)\citenamefont
  {Bradbury}, \citenamefont {Frostig}, \citenamefont {Hawkins}, \citenamefont
  {Johnson}, \citenamefont {Leary}, \citenamefont {Maclaurin}, \citenamefont
  {Necula}, \citenamefont {Paszke}, \citenamefont {Vander{P}las}, \citenamefont
  {Wanderman-{M}ilne},\ and\ \citenamefont {Zhang}}]{Jax2018}%
  \BibitemOpen
  \bibfield  {author} {\bibinfo {author} {\bibfnamefont {J.}~\bibnamefont
  {Bradbury}}, \bibinfo {author} {\bibfnamefont {R.}~\bibnamefont {Frostig}},
  \bibinfo {author} {\bibfnamefont {P.}~\bibnamefont {Hawkins}}, \bibinfo
  {author} {\bibfnamefont {M.~J.}\ \bibnamefont {Johnson}}, \bibinfo {author}
  {\bibfnamefont {C.}~\bibnamefont {Leary}}, \bibinfo {author} {\bibfnamefont
  {D.}~\bibnamefont {Maclaurin}}, \bibinfo {author} {\bibfnamefont
  {G.}~\bibnamefont {Necula}}, \bibinfo {author} {\bibfnamefont
  {A.}~\bibnamefont {Paszke}}, \bibinfo {author} {\bibfnamefont
  {J.}~\bibnamefont {Vander{P}las}}, \bibinfo {author} {\bibfnamefont
  {S.}~\bibnamefont {Wanderman-{M}ilne}}, \ and\ \bibinfo {author}
  {\bibfnamefont {Q.}~\bibnamefont {Zhang}},\ }\href
  {http://github.com/google/jax} {\enquote {\bibinfo {title} {{JAX}: composable
  transformations of {P}ython+{N}um{P}y programs},}\ } (\bibinfo {year}
  {2018})\BibitemShut {NoStop}%
\bibitem [{hig()}]{high_trotter}%
  \BibitemOpen
  \href@noop {} {}\bibinfo {note} {We should also note that suppressing errors
  beyond the third order using the discrete Trotter scheme is not possible due
  to the non-existence of positive decomposition as shown in
  Ref.~\cite{Suzuki1991}. In other words, higher order suppression require
  changing the sign of $\hat H'_{\dr}$, and that is not possible since the sign
  of the nonlinearity $\chi$ is fixed in a device.}\BibitemShut {Stop}%
\bibitem [{\citenamefont {Sitzmann}\ \emph {et~al.}(2020)\citenamefont
  {Sitzmann}, \citenamefont {Martel}, \citenamefont {Bergman}, \citenamefont
  {Lindell},\ and\ \citenamefont {Wetzstein}}]{sitzmann2020implicit}%
  \BibitemOpen
  \bibfield  {author} {\bibinfo {author} {\bibfnamefont {V.}~\bibnamefont
  {Sitzmann}}, \bibinfo {author} {\bibfnamefont {J.}~\bibnamefont {Martel}},
  \bibinfo {author} {\bibfnamefont {A.}~\bibnamefont {Bergman}}, \bibinfo
  {author} {\bibfnamefont {D.}~\bibnamefont {Lindell}}, \ and\ \bibinfo
  {author} {\bibfnamefont {G.}~\bibnamefont {Wetzstein}},\ }\href
  {https://proceedings.neurips.cc/paper_files/paper/2020/file/53c04118df112c13a8c34b38343b9c10-Paper.pdf}
  {\bibfield  {journal} {\bibinfo  {journal} {Advances in Neural Information
  Processing Systems}\ }\textbf {\bibinfo {volume} {33}},\ \bibinfo {pages}
  {7462} (\bibinfo {year} {2020})}\BibitemShut {NoStop}%
\bibitem [{\citenamefont {Suzuki}(1991)}]{Suzuki1991}%
  \BibitemOpen
  \bibfield  {author} {\bibinfo {author} {\bibfnamefont {M.}~\bibnamefont
  {Suzuki}},\ }\href {\doibase 10.1063/1.529425} {\bibfield  {journal}
  {\bibinfo  {journal} {Journal of Mathematical Physics}\ }\textbf {\bibinfo
  {volume} {32}},\ \bibinfo {pages} {400} (\bibinfo {year} {1991})}\BibitemShut
  {NoStop}%
\bibitem [{\citenamefont {Suzuki}(1985)}]{Suzuki1985}%
  \BibitemOpen
  \bibfield  {author} {\bibinfo {author} {\bibfnamefont {M.}~\bibnamefont
  {Suzuki}},\ }\href {\doibase 10.1063/1.526596} {\bibfield  {journal}
  {\bibinfo  {journal} {Journal of Mathematical Physics}\ }\textbf {\bibinfo
  {volume} {26}},\ \bibinfo {pages} {601} (\bibinfo {year} {1985})}\BibitemShut
  {NoStop}%
\bibitem [{\citenamefont {Hoff}\ \emph {et~al.}(2015)\citenamefont {Hoff},
  \citenamefont {Nielsen},\ and\ \citenamefont {Andersen}}]{Hoff2015}%
  \BibitemOpen
  \bibfield  {author} {\bibinfo {author} {\bibfnamefont {U.~B.}\ \bibnamefont
  {Hoff}}, \bibinfo {author} {\bibfnamefont {B.~M.}\ \bibnamefont {Nielsen}}, \
  and\ \bibinfo {author} {\bibfnamefont {U.~L.}\ \bibnamefont {Andersen}},\
  }\href {\doibase 10.1364/OE.23.012013} {\bibfield  {journal} {\bibinfo
  {journal} {Optics Express}\ }\textbf {\bibinfo {volume} {23}},\ \bibinfo
  {pages} {12013} (\bibinfo {year} {2015})}\BibitemShut {NoStop}%
\bibitem [{\citenamefont {Mittal}\ \emph {et~al.}(2021)\citenamefont {Mittal},
  \citenamefont {Moille}, \citenamefont {Srinivasan}, \citenamefont {Chembo},\
  and\ \citenamefont {Hafezi}}]{Mittal2021}%
  \BibitemOpen
  \bibfield  {author} {\bibinfo {author} {\bibfnamefont {S.}~\bibnamefont
  {Mittal}}, \bibinfo {author} {\bibfnamefont {G.}~\bibnamefont {Moille}},
  \bibinfo {author} {\bibfnamefont {K.}~\bibnamefont {Srinivasan}}, \bibinfo
  {author} {\bibfnamefont {Y.~K.}\ \bibnamefont {Chembo}}, \ and\ \bibinfo
  {author} {\bibfnamefont {M.}~\bibnamefont {Hafezi}},\ }\href {\doibase
  10.1038/s41567-021-01302-3} {\bibfield  {journal} {\bibinfo  {journal}
  {Nature Physics}\ }\textbf {\bibinfo {volume} {17}},\ \bibinfo {pages} {1169}
  (\bibinfo {year} {2021})}\BibitemShut {NoStop}%
\bibitem [{\citenamefont {Vernon}\ \emph {et~al.}(2019)\citenamefont {Vernon},
  \citenamefont {Quesada}, \citenamefont {Liscidini}, \citenamefont {Morrison},
  \citenamefont {Menotti}, \citenamefont {Tan},\ and\ \citenamefont
  {Sipe}}]{Vernon2019}%
  \BibitemOpen
  \bibfield  {author} {\bibinfo {author} {\bibfnamefont {Z.}~\bibnamefont
  {Vernon}}, \bibinfo {author} {\bibfnamefont {N.}~\bibnamefont {Quesada}},
  \bibinfo {author} {\bibfnamefont {M.}~\bibnamefont {Liscidini}}, \bibinfo
  {author} {\bibfnamefont {B.}~\bibnamefont {Morrison}}, \bibinfo {author}
  {\bibfnamefont {M.}~\bibnamefont {Menotti}}, \bibinfo {author} {\bibfnamefont
  {K.}~\bibnamefont {Tan}}, \ and\ \bibinfo {author} {\bibfnamefont
  {J.}~\bibnamefont {Sipe}},\ }\href {\doibase
  10.1103/PhysRevApplied.12.064024} {\bibfield  {journal} {\bibinfo  {journal}
  {Physical Review Applied}\ }\textbf {\bibinfo {volume} {12}},\ \bibinfo
  {pages} {064024} (\bibinfo {year} {2019})}\BibitemShut {NoStop}%
\end{thebibliography}%


\begin{thebibliography}{10}%
\makeatletter
\providecommand \@ifxundefined [1]{%
 \@ifx{#1\undefined}
}%
\providecommand \@ifnum [1]{%
 \ifnum #1\expandafter \@firstoftwo
 \else \expandafter \@secondoftwo
 \fi
}%
\providecommand \@ifx [1]{%
 \ifx #1\expandafter \@firstoftwo
 \else \expandafter \@secondoftwo
 \fi
}%
\providecommand \natexlab [1]{#1}%
\providecommand \enquote  [1]{``#1''}%
\providecommand \bibnamefont  [1]{#1}%
\providecommand \bibfnamefont [1]{#1}%
\providecommand \citenamefont [1]{#1}%
\providecommand \href@noop [0]{\@secondoftwo}%
\providecommand \href [0]{\begingroup \@sanitize@url \@href}%
\providecommand \@href[1]{\@@startlink{#1}\@@href}%
\providecommand \@@href[1]{\endgroup#1\@@endlink}%
\providecommand \@sanitize@url [0]{\catcode `\\12\catcode `\$12\catcode
  `\&12\catcode `\#12\catcode `\^12\catcode `\_12\catcode `\%12\relax}%
\providecommand \@@startlink[1]{}%
\providecommand \@@endlink[0]{}%
\providecommand \url  [0]{\begingroup\@sanitize@url \@url }%
\providecommand \@url [1]{\endgroup\@href {#1}{\urlprefix }}%
\providecommand \urlprefix  [0]{URL }%
\providecommand \Eprint [0]{\href }%
\providecommand \doibase [0]{http://dx.doi.org/}%
\providecommand \selectlanguage [0]{\@gobble}%
\providecommand \bibinfo  [0]{\@secondoftwo}%
\providecommand \bibfield  [0]{\@secondoftwo}%
\providecommand \translation [1]{[#1]}%
\providecommand \BibitemOpen [0]{}%
\providecommand \bibitemStop [0]{}%
\providecommand \bibitemNoStop [0]{.\EOS\space}%
\providecommand \EOS [0]{\spacefactor3000\relax}%
\providecommand \BibitemShut  [1]{\csname bibitem#1\endcsname}%
\let\auto@bib@innerbib\@empty
\bibitem [{\citenamefont {Boscain}\ \emph {et~al.}(2021)\citenamefont
  {Boscain}, \citenamefont {Sigalotti},\ and\ \citenamefont
  {Sugny}}]{boscain2021introSM}%
  \BibitemOpen
  \bibfield  {author} {\bibinfo {author} {\bibfnamefont {U.}~\bibnamefont
  {Boscain}}, \bibinfo {author} {\bibfnamefont {M.}~\bibnamefont {Sigalotti}},
  \ and\ \bibinfo {author} {\bibfnamefont {D.}~\bibnamefont {Sugny}},\ }\href
  {\doibase 10.1103/PRXQuantum.2.030203} {\bibfield  {journal} {\bibinfo
  {journal} {PRX Quantum}\ }\textbf {\bibinfo {volume} {2}},\ \bibinfo {pages}
  {030203} (\bibinfo {year} {2021})}\BibitemShut {NoStop}%
\bibitem [{\citenamefont {Krastanov}\ \emph {et~al.}(2015)\citenamefont
  {Krastanov}, \citenamefont {Albert}, \citenamefont {Shen}, \citenamefont
  {Zou}, \citenamefont {Heeres}, \citenamefont {Vlastakis}, \citenamefont
  {Schoelkopf},\ and\ \citenamefont {Jiang}}]{Krastanov2015SM}%
  \BibitemOpen
  \bibfield  {author} {\bibinfo {author} {\bibfnamefont {S.}~\bibnamefont
  {Krastanov}}, \bibinfo {author} {\bibfnamefont {V.~V.}\ \bibnamefont
  {Albert}}, \bibinfo {author} {\bibfnamefont {C.}~\bibnamefont {Shen}},
  \bibinfo {author} {\bibfnamefont {C.-L.}\ \bibnamefont {Zou}}, \bibinfo
  {author} {\bibfnamefont {R.~W.}\ \bibnamefont {Heeres}}, \bibinfo {author}
  {\bibfnamefont {B.}~\bibnamefont {Vlastakis}}, \bibinfo {author}
  {\bibfnamefont {R.~J.}\ \bibnamefont {Schoelkopf}}, \ and\ \bibinfo {author}
  {\bibfnamefont {L.}~\bibnamefont {Jiang}},\ }\href {\doibase
  10.1103/physreva.92.040303} {\bibfield  {journal} {\bibinfo  {journal}
  {Physical Review A}\ }\textbf {\bibinfo {volume} {92}},\ \bibinfo {pages}
  {040303(R)} (\bibinfo {year} {2015})}\BibitemShut {NoStop}%
\bibitem [{\citenamefont {Schirmer}\ \emph {et~al.}(2001)\citenamefont
  {Schirmer}, \citenamefont {Fu},\ and\ \citenamefont
  {Solomon}}]{Schirmer2001SM}%
  \BibitemOpen
  \bibfield  {author} {\bibinfo {author} {\bibfnamefont {S.~G.}\ \bibnamefont
  {Schirmer}}, \bibinfo {author} {\bibfnamefont {H.}~\bibnamefont {Fu}}, \ and\
  \bibinfo {author} {\bibfnamefont {A.~I.}\ \bibnamefont {Solomon}},\ }\href
  {\doibase 10.1103/physreva.63.063410} {\bibfield  {journal} {\bibinfo
  {journal} {Physical Review A}\ }\textbf {\bibinfo {volume} {63}},\ \bibinfo
  {pages} {063410} (\bibinfo {year} {2001})}\BibitemShut {NoStop}%
\bibitem [{\citenamefont {Lingenfelter}\ \emph {et~al.}(2021)\citenamefont
  {Lingenfelter}, \citenamefont {Roberts},\ and\ \citenamefont
  {Clerk}}]{Lingenfelter2021SM}%
  \BibitemOpen
  \bibfield  {author} {\bibinfo {author} {\bibfnamefont {A.}~\bibnamefont
  {Lingenfelter}}, \bibinfo {author} {\bibfnamefont {D.}~\bibnamefont
  {Roberts}}, \ and\ \bibinfo {author} {\bibfnamefont {A.~A.}\ \bibnamefont
  {Clerk}},\ }\href {\doibase 10.1126/sciadv.abj1916} {\bibfield  {journal}
  {\bibinfo  {journal} {Science Advances}\ }\textbf {\bibinfo {volume} {7}},\
  \bibinfo {pages} {eabj1916} (\bibinfo {year} {2021})}\BibitemShut {NoStop}%
\bibitem [{\citenamefont {Suzuki}(1985)}]{Suzuki1985SM}%
  \BibitemOpen
  \bibfield  {author} {\bibinfo {author} {\bibfnamefont {M.}~\bibnamefont
  {Suzuki}},\ }\href {\doibase 10.1063/1.526596} {\bibfield  {journal}
  {\bibinfo  {journal} {Journal of Mathematical Physics}\ }\textbf {\bibinfo
  {volume} {26}},\ \bibinfo {pages} {601} (\bibinfo {year} {1985})}\BibitemShut
  {NoStop}%
\bibitem [{\citenamefont {Choi}\ \emph {et~al.}(2017)\citenamefont {Choi},
  \citenamefont {Heuck},\ and\ \citenamefont {Englund}}]{Choi2017SM}%
  \BibitemOpen
  \bibfield  {author} {\bibinfo {author} {\bibfnamefont {H.}~\bibnamefont
  {Choi}}, \bibinfo {author} {\bibfnamefont {M.}~\bibnamefont {Heuck}}, \ and\
  \bibinfo {author} {\bibfnamefont {D.}~\bibnamefont {Englund}},\ }\href
  {\doibase 10.1103/PhysRevLett.118.223605} {\bibfield  {journal} {\bibinfo
  {journal} {Physical Review Letters}\ }\textbf {\bibinfo {volume} {118}},\
  \bibinfo {pages} {223605} (\bibinfo {year} {2017})}\BibitemShut {NoStop}%
\bibitem [{\citenamefont {Hoff}\ \emph {et~al.}(2015)\citenamefont {Hoff},
  \citenamefont {Nielsen},\ and\ \citenamefont {Andersen}}]{Hoff2015SM}%
  \BibitemOpen
  \bibfield  {author} {\bibinfo {author} {\bibfnamefont {U.~B.}\ \bibnamefont
  {Hoff}}, \bibinfo {author} {\bibfnamefont {B.~M.}\ \bibnamefont {Nielsen}}, \
  and\ \bibinfo {author} {\bibfnamefont {U.~L.}\ \bibnamefont {Andersen}},\
  }\href {\doibase 10.1364/OE.23.012013} {\bibfield  {journal} {\bibinfo
  {journal} {Optics Express}\ }\textbf {\bibinfo {volume} {23}},\ \bibinfo
  {pages} {12013} (\bibinfo {year} {2015})}\BibitemShut {NoStop}%
\bibitem [{\citenamefont {Mittal}\ \emph {et~al.}(2021)\citenamefont {Mittal},
  \citenamefont {Moille}, \citenamefont {Srinivasan}, \citenamefont {Chembo},\
  and\ \citenamefont {Hafezi}}]{Mittal2021SM}%
  \BibitemOpen
  \bibfield  {author} {\bibinfo {author} {\bibfnamefont {S.}~\bibnamefont
  {Mittal}}, \bibinfo {author} {\bibfnamefont {G.}~\bibnamefont {Moille}},
  \bibinfo {author} {\bibfnamefont {K.}~\bibnamefont {Srinivasan}}, \bibinfo
  {author} {\bibfnamefont {Y.~K.}\ \bibnamefont {Chembo}}, \ and\ \bibinfo
  {author} {\bibfnamefont {M.}~\bibnamefont {Hafezi}},\ }\href {\doibase
  10.1038/s41567-021-01302-3} {\bibfield  {journal} {\bibinfo  {journal}
  {Nature Physics}\ }\textbf {\bibinfo {volume} {17}},\ \bibinfo {pages} {1169}
  (\bibinfo {year} {2021})}\BibitemShut {NoStop}%
\bibitem [{\citenamefont {Vernon}\ \emph {et~al.}(2019)\citenamefont {Vernon},
  \citenamefont {Quesada}, \citenamefont {Liscidini}, \citenamefont {Morrison},
  \citenamefont {Menotti}, \citenamefont {Tan},\ and\ \citenamefont
  {Sipe}}]{Vernon2019SM}%
  \BibitemOpen
  \bibfield  {author} {\bibinfo {author} {\bibfnamefont {Z.}~\bibnamefont
  {Vernon}}, \bibinfo {author} {\bibfnamefont {N.}~\bibnamefont {Quesada}},
  \bibinfo {author} {\bibfnamefont {M.}~\bibnamefont {Liscidini}}, \bibinfo
  {author} {\bibfnamefont {B.}~\bibnamefont {Morrison}}, \bibinfo {author}
  {\bibfnamefont {M.}~\bibnamefont {Menotti}}, \bibinfo {author} {\bibfnamefont
  {K.}~\bibnamefont {Tan}}, \ and\ \bibinfo {author} {\bibfnamefont
  {J.}~\bibnamefont {Sipe}},\ }\href {\doibase
  10.1103/PhysRevApplied.12.064024} {\bibfield  {journal} {\bibinfo  {journal}
  {Physical Review Applied}\ }\textbf {\bibinfo {volume} {12}},\ \bibinfo
  {pages} {064024} (\bibinfo {year} {2019})}\BibitemShut {NoStop}%
\bibitem [{\citenamefont {Sitzmann}\ \emph {et~al.}(2020)\citenamefont
  {Sitzmann}, \citenamefont {Martel}, \citenamefont {Bergman}, \citenamefont
  {Lindell},\ and\ \citenamefont {Wetzstein}}]{sitzmann2020implicitSM}%
  \BibitemOpen
  \bibfield  {author} {\bibinfo {author} {\bibfnamefont {V.}~\bibnamefont
  {Sitzmann}}, \bibinfo {author} {\bibfnamefont {J.}~\bibnamefont {Martel}},
  \bibinfo {author} {\bibfnamefont {A.}~\bibnamefont {Bergman}}, \bibinfo
  {author} {\bibfnamefont {D.}~\bibnamefont {Lindell}}, \ and\ \bibinfo
  {author} {\bibfnamefont {G.}~\bibnamefont {Wetzstein}},\ }\href
  {https://proceedings.neurips.cc/paper_files/paper/2020/file/53c04118df112c13a8c34b38343b9c10-Paper.pdf}
  {\bibfield  {journal} {\bibinfo  {journal} {Advances in Neural Information
  Processing Systems}\ }\textbf {\bibinfo {volume} {33}},\ \bibinfo {pages}
  {7462} (\bibinfo {year} {2020})}\BibitemShut {NoStop}%
\end{thebibliography}%

\end{document}


\title{Supplemental Material for ``Universal Control in Bosonic Systems with Weak Kerr Nonlinearities"}

\author{Ming Yuan}
\email{yuanming@uchicago.edu}
\affiliation{Pritzker School of Molecular Engineering, The University of Chicago, Chicago, IL 60637, USA}
\author{Alireza Seif}
\affiliation{Pritzker School of Molecular Engineering, The University of Chicago, Chicago, IL 60637, USA}
\author{Andrew Lingenfelter}
\affiliation{Department of Physics, The University of Chicago, Chicago, IL 60637, USA}

\author{David I. Schuster}
\affiliation{Pritzker School of Molecular Engineering, The University of Chicago, Chicago, IL 60637, USA}
\affiliation{Department of Physics, The University of Chicago, Chicago, IL 60637, USA}
\affiliation{James Franck Institute, The University of Chicago, Chicago, IL 60637, USA}

\author{Aashish A. Clerk}
\affiliation{Pritzker School of Molecular Engineering, The University of Chicago, Chicago, IL 60637, USA}
\author{Liang Jiang}
\email{liangjiang@uchicago.edu}
\affiliation{Pritzker School of Molecular Engineering, The University of Chicago, Chicago, IL 60637, USA}

\date{\today}

\maketitle

\section{Detailed universality proof}\label{app:univ_pf}

In this part, we show that the Hamiltonian $\hat H_\dr$ (see Eq.~\eqref{Hdr}) can generate arbitrary unitary operations in U($N$) within the $N = r+1$ dimensional blockade subspace $\mathcal{H}_b$. Since $r$ is an adjustable parameter, our protocol allows for ``universal control" in bosonic systems with Kerr nonlinearities in the sense that any unitary operator with any chosen dimension is realizable in the system.

In quantum control theory~\cite{boscain2021introSM}, a Hamiltonian defined in an $N$ dimensional Hilbert space spanned by $\{\ket{k}\}_{k=0}^{N-1}$ contains a drift part $\hat H_d$ and several control parts $\hat H_{c, j}$. It can be expressed as
\begin{equation}\label{Ht}
    \hat H(t) = \hat H_d + \sum_j v_j(t) \hat H_{c, j}.
\end{equation}
Here, we consider the drift and one of the control parts with the following form 
\begin{equation}\label{HdHc}
    \begin{split}
        & \hat H_d = \sum_{k=0}^{N-1} E_k \ketbra{k},\\
        & \hat H_{c, 1} = \sum_{k=0}^{N-2} d_k (\ketbra{k+1}{k} + \ketbra{k}{k+1} ).
    \end{split}
\end{equation}
Here $d_k \in \mathbb{R}$ and $d_k \neq 0$. The ``universal control"~\cite{Krastanov2015SM} is named as the ability to realize any unitary operation $\hat U_\tar$ in $U(N)$ with properly chosen $v_j(t)$ and evolution time $T$, such that
\begin{equation}
    \hat U_\tar = \hat U(T) = \mathcal{T}\exp[-i\int_0^T \hat H(t') \dif t'].
\end{equation}

A Theorem in Ref.~\cite{Schirmer2001SM} suggests a sufficient condition for the choice of $E_k$ and $d_k$ to make the system universally controllable. We first repeat the theorem here and check that our Hamiltonian $\hat H_\dr$ restricted in blockade subspace ($\hat\Pi_r \hat H_\dr \hat\Pi_r$, see Eq.~\eqref{eq:ham_proj}) satisfies those criteria.

\begin{thm}[Ref.~\cite{Schirmer2001SM}]\label{ThmUniversal}
Denote $\mu_k = E_k - E_{k+1}$. If $\mu_{0} \neq 0$ and $\mu_k^2 \neq \mu_{0}^2$ for $k > 0$ (or similarly if $\mu_{N-2} \neq 0$ and $\mu_k^2 \neq \mu_{N-2}^2$ for $k < N-2$), then the dynamical Lie group of the system $\hat H(t)$ defined in Eq. (\ref{Ht}, \ref{HdHc}) is at least SU($N$). Further, if $\Tr[\hat H_d] \neq 0$, the dynamical Lie group is U($N$).
\end{thm}

We notice that $\hat\Pi_r \hat H_\dr \hat\Pi_r$ contains the drift part $\hat H_{d,0}$ and two control parts $\hat H_{c,R}$ and $\hat H_{c,I}$. It is easy to verify that $\hat H_{c,R}$ meets the requirement for the control part. Then we should check the nonlinear condition for $\hat H_{d,0}$. In this case, we have $E_k = \chi (k^2-k)/2 + \Delta_0 k$, which gives the nearest energy difference $\mu_k = -\chi k - \Delta_0$. To match the condition in Theorem \ref{ThmUniversal}, we need to make sure $\mu_0^2 \neq \mu_{N-2}^2$, which leads to $r \neq -\frac{2\Delta_0}{\chi} + 1$. Further, since $\mu_k$ is monotonic in $k$, we cannot find both $k_1, k_2$ ($0 < k_1, k_2 < N-2$) such that $\mu_0^2 = \mu_{k_1}^2$ and $\mu_{N-2}^2 = \mu_{k_2}^2$. This concludes the proof of the universal controllability of our system, even if we can fix $\Im[\alpha(t)] = 0$ all the time. 



Moreover, the ability to control $\Im[\alpha(t)]$ provides us with an additional degree of freedom to control the system, which in principle makes it possible to perform operations arbitrarily fast. To demonstrate this, we first show that the nested commutators between $i\hat H_{c,R}$ and $i\hat H_{c,I}$ are sufficient to form a complete basis of the Lie algebra associated with SU($N$). The reason is that $[i\hat H_{c,R}, i\hat H_{c,I}] = \sum_{n=0}^r -2[3n^2 - (4r+1)n + r^2]i \ketbra{n}{n}$ is diagonal in Fock basis, has zero trace, and fulfills the anharmonicity requirement as that for the drift term $\hat H_d$ in Theorem \ref{ThmUniversal}. Further, as proved in Theorem \ref{ThmUniversal}, $[i\hat H_{c,R}, i\hat H_{c,I}] $ and $i\hat H_{c,R}$ are sufficient to generate a set of complete basis of Lie algebra $\mathfrak{su}(N)$. In practice, the overall phase for unitary operations has no physical meaning, which makes SU($N$) group sufficient for universality. Second, since $\chi\Re[\alpha(t)] \hat H_{c,R} + \chi\Im[\alpha(t)] \hat H_{c,I}$ is sufficient to achieve the desired unitary, we can simply increase the amplitude of $\alpha(t)$ to do things arbitrarily fast. In the short time limit, the drift term $\hat H_{d,0}$ does not contribute to the dynamics and therefore does not impose a speed constraint.

\section{Basic error analysis}\label{app:err_ana}

In this section, we analyze the errors in implementing a desired operation using our protocol where only 1-photon drive is allowed. Notice that, even if all the driving parameters are perfectly implemented, there are errors stemming from photon loss and inaccuracies of the Trotter design in Eq.~\eqref{UniApprox}. Reducing the total time of the protocol $T$ helps to mitigate the photon loss errors, but at the same time increases the required input power, which could be limited in practice. This limited power also introduces coherent errors in the Trotter approximation of the unitary operator of interest. Therefore, there is a trade-off between these two sources of error,  which we discuss in detail in the following. 

For simplicity, we only consider the error in state preparation tasks characterized by infidelity 
\begin{equation}\label{infidel}
    \epsilon = 1 - \bra{\psi_\tar}\hat\rho(T)\ket{\psi_\tar}, 
\end{equation}
where $\ket{\psi_\tar}$ is a pure state that we are interested in preparing and $\hat\rho(T)$ is the state of the system at the end of the evolution. Here, $\hat\rho(T)$ is the state obtained after evolving for time $T$ under the Lindblad equation
\begin{equation}\label{eqn:lindblad}
    \frac{\dif \hat \rho}{\dif t} = -i[\hat H'_\dr, \hat{\rho}] + \kappa \mathcal{D}[\hat{a}] \hat\rho,
\end{equation}
where $\hat H'_\dr$ is introduced in Eq.~\eqref{eqn:new_single_drive}.

To study the scaling of error $\epsilon$ with relevant physical parameters in the problem, we focus our attention on the case of single-photon state preparation when the blockade subsystem has only 2 dimension and $\alpha(t)$ is constant, as originally proposed in Ref.~\cite{Lingenfelter2021SM}.  Therefore, in this case we set $\ket{\psi_\tar} = \ket{1}$, and  $\hat{\rho}(0)=\ketbra{0}{0}$. To prepare the target state $\ket{1}$, we choose $\Delta_0 = 0$ and $\alpha(t) = \alpha$, a constant that is assumed to be a real number for simplicity. Therefore, the effective Hamiltonian in the blockade subspace is
\begin{equation}\label{eqn:h_qb_rd}
    \hat H^\qb_\dr = -\chi\alpha \hat \sigma_x. 
\end{equation}
Here $\hat \sigma_x = \ketbra{0}{1} + \ketbra{1}{0}$. Moreover, in this subspace the photon loss dissipator $\mathcal{D}[\hat{a}]$ in Eq.~\eqref{eqn:lindblad} is now simply modified to $\mathcal{D}[\hat\sigma_-]$, where $\hat\sigma_- = \ketbra{0}{1}$. Note that in the ideal case, when there is no loss or Trotter error, we can perfectly prepare $\ket{1}$ by evolving $\ket{0}$ under $\hat H^\qb_\dr$ in Eq.~\eqref{eqn:h_qb_rd} for time $T = \frac{\pi}{2\chi\alpha}$. 
For simplicity here we ignore the part of the dynamics where $\alpha(t)$ increases from $0$ to $\alpha$ in the beginning and decreases to $0$ in the end~\cite{Lingenfelter2021SM}, while only discussing errors with a constant $\alpha(t)$. In fact, using a time-dependent $\alpha(t)$ will introduce additional imperfection for Trotter approximation. However, as discussed in Sec.~\ref{AppTimeDep}, given the assumption that $\alpha(t)$ varies slowly over time, these additional errors are higher-order effects compared with the major Trotter error scaling we focus on later. 

To find the scaling properties of $\epsilon$, we treat the photon loss ($\epsilon_{\rm{loss}}$)  and Trotter ($\epsilon_{\rm{tt}}$) errors independently.

First, for the photon loss error, we assume that the Hamiltonian $\hat H^\qb_\dr$ in Eq.~\eqref{eqn:h_qb_rd} is perfectly implemented so that the quantum state is confined to the blockade subspace all the time. If $\chi\alpha \gg \kappa$, we can treat the dissipative term $\kappa \mathcal{D}[\sigma_-]$ perturbatively.  So, to the lowest order, the error from the photon loss process satisfies $\epsilon_\loss = c_1 \kappa T$, where the coefficient $c_1$ in general depends on the function $\alpha(t)$. In our case, we find that $c_1 = \frac{3}{8}$ (see Sec.~\ref{app:esti_coe}). 

Next, to estimate the error from the Trotter approximation, we ignore the photon loss process and assume that the system undergoes unitary evolution with Hamiltonian $\hat H'_\dr$ shown in Eq.~\eqref{eqn:new_single_drive}. For simplicity, we use the discrete Trotter formula to illustrate analysis, but the results apply to the proposed continuous version as well (see Sec.~\ref{app:misc}). Let $\hat U_\tar = e^{-i\frac{\hat H_1 + \hat H_2}{2} \delta T}$ and $\hat U = e^{-i\hat H_1\delta T/4} e^{-i\hat H_2\delta T/2} e^{-i\hat H_1\delta T/4}$. Using the discrete version of Trotter formula~\cite{Suzuki1985SM}, we have
\begin{equation}\label{eqn:coh_dev}
    \hat U_\tar^{-1/2} \hat U \hat U_\tar^{-1/2} = \exp\{(\delta T)^3 \hat A_3 + O[(\delta T)^5]\},
\end{equation}
where $\hat A_3 = i[[\hat H_1, \hat H_2], \hat H_1 + 2\hat H_2]/192$ is anti-Hermitian. Further, recall the form of $\hat H'_\dr[\tilde \alpha]$. Since we work in the regime that $|\alpha| \gg \sqrt{N}$ ($N$ is the dimension of the blockade subspace), the dominant contribution to $\hat A_3$ comes from the $(\frac{\chi}{2} \alpha^2_{1,2} \hat a^{\da 2} + \hc)$ terms in the Hamiltonian $\hat H_1$ and $\hat H_2$, and therefore the matrix elements of $\hat A_3$ that we focus on scale as $O[(\chi\alpha^2)^3]$. Finally, as mentioned before, we need to sequentially apply $M = T/\delta T$ repetitions of the Trotter operation $\hat U$ to approximately achieve our desired unitary $\hat U_\tar^M$, where $\ketbra{\psi_\tar} = \hat U_\tar^M \hat\rho(0) \hat U_\tar^{\da M}$. Therefore, we find
\begin{equation}
\begin{split}
    \epsilon_{\ttt} &= 1 - \bra{\psi_\tar} \hat U^M \hat \rho(0)\hat U^{\da M}\ket{\psi_\tar}\\
    &= O[(M (\chi\alpha^2\delta T)^3)^2 ].
\end{split}
\end{equation}
Recalling the relation that $T \propto 1/(\chi\alpha)$ and $M = T/\delta T$, we finally obtain $\epsilon_\ttt = c_2 / [M^4 (\chi T)^6]$, where $c_2$ is again the coefficient that depends on specific form of our pulse design and can be estimated numerically (see Sec.~\ref{app:esti_coe}). Note that in our numerical work $f(t)$ function is chosen as
\begin{equation}\label{SemiRot}
    f(t) = \frac{i \pi}{2} e^{-i 2\pi |s(t) - 1/2|},
\end{equation}
where $s(t) = t/\delta T - \lfloor t/\delta T\rfloor$. This choice of $f(t)$ also satisfies the requirements shown in Eq.~(\ref{f_even}, \ref{f_avg}), and also $|f(t)|$ is constant over time. In this situation, the reference displaced frame (characterized by $\tilde\alpha(t)$) oscillates back and forth along a semi-circle. In practice, it can be implemented with 1-photon drives using two interleaved tones with frequencies $\omega_1 \pm \omega_r$.

Since errors from the loss and Trotter approximation are independent, to the lowest order they can be added directly to find the total infidelity as
\begin{equation}\label{errtot}
    \epsilon_\tot = \epsilon_\loss + \epsilon_\ttt = c_1 \kappa T + \frac{c_2}{M^4 (\chi T)^6}.
\end{equation}

If the driving amplitude is unbounded, we can use an arbitrarily short time $T$ with arbitrary fast oscillating $\tilde \alpha(t)$ ($M \to \infty$) such that both $\epsilon_\loss$ and $\epsilon_\ttt$ are suppressed to zero. However, in practice there are possible limitations that prevent us from doing that. Here we consider a key limitation, namely the input power $P_\inp$ that can be applied to the system and evaluate the optimal $\epsilon_\tot$ that we can achieve under this constraint. 

\begin{figure}[t!]
\centering
\includegraphics[scale = 0.49]{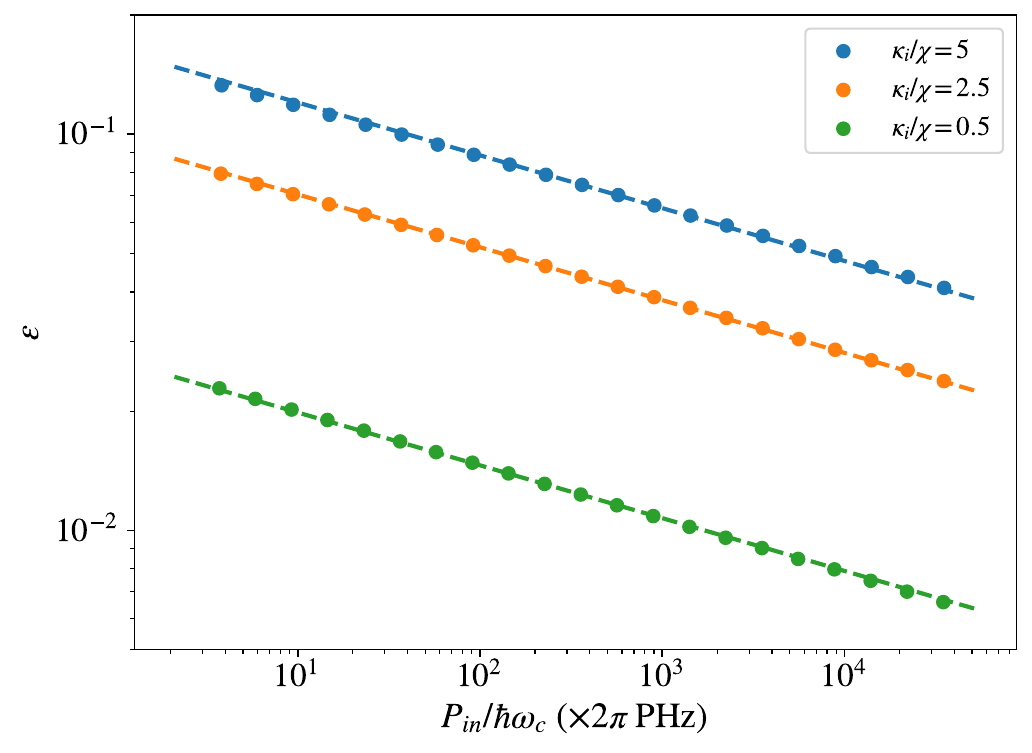}
\caption{Numerical results of the infidelity of Fock state $\ket{1}$ preparation with given input power and corresponding optimal choice of gate time. The choice of Kerr parameter is $\chi = 2\pi \times 3$ kHz. The dynamics is calculated with the assumption that RWA is valid. Different colors indicate different $\kappa_i/\chi$ ratio. The dashed lines are references for $\epsilon \propto P_\inp^{-2/15}$ scaling.}
\label{fig:Power}
\end{figure}

First, we assume that $\chi|\alpha|^2 \ll \omega_c$, which can be checked self-consistently since otherwise RWA will not apply, as discussed in Sec.~\ref{app:self_con}. This assumption allows us approximately write down the input power as 
\begin{equation}
    P_\inp \simeq \frac{|\Lambda_1|^2}{\kappa_e} \hbar\omega_c,
\end{equation}
where $\kappa_e$ is the external loss rate induced by coupling to the driving port. The total loss $\kappa$ is composed of $\kappa = \kappa_e + \kappa_i$, where $\kappa_i$ is the internal loss rate of the resonator. We notice from Eq.~\eqref{eq:Lambda1_tilde} that, $\Lambda_1 \sim O(\chi\tilde\alpha^3) + i\dot {\tilde \alpha}$. We then need to estimate the contribution from $\dot {\tilde\alpha}$. In our design we have $\dot {\tilde\alpha} \sim O(\alpha M/T)$. Besides, since we are only considering lowest order errors from the Trotter expansion, the Trotter error should be small. Therefore, we should have $\chi\alpha^2 \delta T \ll 1$, or equivalently $\chi\alpha^2 \ll \omega_r$. This implies that $M \gg \chi\alpha^2 T$, and as a result $i\dot{\tilde \alpha}$ is the dominating term in $\Lambda_1$. So, we can write $P_\inp$ as
\begin{equation}\label{Power}
    P_\inp \sim O(\frac{M^2 \alpha^2}{\kappa_e T^2}) :=\frac{c_3 M^2}{\kappa_e \chi^2 T^4 },
\end{equation}
where $\hbar\omega_c$ and the proportionality constant originating from $f(t)$ are both absorbed in $c_3$. By eliminating $M$ in Eq.~\eqref{errtot} using Eq.~\eqref{Power} we obtain
\begin{equation}\label{EPSITOT}
    \epsilon_\tot = c_1 (\kappa_e + \kappa_i) T + \frac{c_2 c_3^2}{P_\inp^2 \kappa_e^2 \chi^{10} T^{14}}.
\end{equation}
As a result, we find that the scaling of $T^\opt$, namely the optimal time that minimizes $\epsilon_\tot$, and the corresponding optimal error $\epsilon_\tot^\opt$ is given by
\begin{equation}\label{opti}
    \begin{split}
        T^\opt &= \left(\frac{14c_2 c_3^2}{c_1}\right)^{1/15}\cdot\frac{1}{P_\inp^{2/15} \chi^{2/3}\kappa_e^{2/15} (\kappa_e + \kappa_i)^{1/15} },\\
        \epsilon^\opt_\tot &= 15\left(\frac{c_1^{14}c_2 c_3^2}{14^{14}}\right)^{1/15} \cdot \left(\frac{1}{P_\inp^{2/15} \chi^{2/3}}\frac{(\kappa_e + \kappa_i)^{14/15}}{\kappa_e^{2/15}}\right).
    \end{split}
\end{equation}

It is also natural to assume that $\kappa_i$ is a fixed property of the device while $\kappa_e$ is adjustable to further optimize $\epsilon^\opt_\tot$. We can find the optimal ratio of $\kappa_e/ \kappa_i =\frac{1}{6}$ that minimizes $\epsilon^\opt_\tot$  while keeping $\chi$, $\kappa_i$ and input power $P_\inp$ fixed and using the optimal $T^\opt$. This gives
\begin{equation}
    \epsilon^\opt_\tot = \frac{15(c_1^{14}c_2 c_3^2)^{1/15}}{2^{14/15}\cdot 6^{4/5}} \cdot \left(\frac{\kappa_i^{4/5}}{P_\inp^{2/15} \chi^{2/3}}\right).
\end{equation}

We also numerically investigate the infidelity $\epsilon$ as prescribed by Eq.~(\ref{infidel}) as a function of input power. To find the optimal time for the protocol in our numerical simulations we first need to determine the constants in Eq.~\eqref{EPSITOT}. More details can be found in Sec.~\ref{app:esti_coe}. We first obtain $c_2$ by varying $M$ and $T$ in the simulations and extracting the corresponding proportionality constant in Eq.~\eqref{errtot}. We then obtain $c_3$ by using the explicit form of $f(t)$ in Eq.~(\ref{SemiRot}), which results in  $c_3 \simeq \frac{\pi^6}{4} \hbar \omega_c$. Therefore, together with $c_1 = \frac{3}{8}$ from perturbative analysis, we can then find $T^\opt$ for a given power, which subsequently determines the required $M$ using Eq.~\eqref{Power}. In this way, both the optimal protocol time $T^\opt$ and $\tilde \alpha(t)$ are fully determined. Here we also use the optimal choice of $\kappa_e = \frac{1}{6} \kappa_i$.

In FIG.~\ref{fig:Power}, we observe that the numerically obtained infidelities agree well with the estimated $\epsilon \propto P_\inp^{-2/15}$ scaling in a wide range of $P_\inp$ values.  

Finally, in the numerical simulation we have assumed the validity of RWA so that the dynamics is irrelevant with $\omega_c$ after going to the rotating frame. We notice that if the required $\alpha$ or $M$ (equivalently $\omega_r$) is so large such that $\Lambda_1$ is comparable with the frequency of the resonator $\omega_c$, then those off-resonant terms which has been ignored under RWA in the beginning may lead to non-negligible effects on the dynamics. In Sec.~\ref{app:self_con}, we use scaling analysis to briefly discuss those effects and strategies to partially compensate them by adjusting $\Lambda_1(t)$ and $\omega_1(t)$, or driving both charge and flux quadratures together rather than only one of them on hardware.



\section{Trotter errors with time-dependent $\alpha(t)$}\label{AppTimeDep}

In this section, we will also discuss the error scaling properties of our protocol, but specifically focus on the Trotter error in the case that $\alpha(t)$ designed by optimal control algorithm is a slowly varied function that depends on time $t$. Here, by ``slowly varied", we mean the time derivative of $\alpha(t)$ scales as $\dot\alpha(t) \sim O(\alpha/T)$, and in general $\dif{}^{n} \alpha(t) / \dif t^n \sim O(\alpha/T^n)$. To be more precise, we will calculate the difference between both sides of Eq.~(\ref{UniApprox}) in detail, where on the left-hand side is the evolution we can achieve with only 1-photon drive, and on the right-hand side is the target unitary operation that we want to achieve. Notice that, unlike the constant $\alpha$ case in the main text that $T \propto 1/(\chi\alpha)$, here we do not have such property rigorously. But, it is still reasonable to assume that $T$ roughly scales as $O(1/\chi\alpha)$ since the dominant part in $\hat H_\dr$ is still $[\chi\alpha(t)\hat a^\da (\hat n - r) + \hc]$ if $\alpha$ is large. We will keep this assumption in the following derivation. 

First, let us calculate the difference of unitary operation within each time slice $t\in [k\delta T, (k+1)\delta T]$ where $k$ is an integer and $0\leq k < M$. We denote $\hat U_k = \mathcal{T}\exp\{-i \int_{k\delta T}^{(k+1)\delta T} \hat H'_\dr[\tilde \alpha(t)] \dif t\}$ and $\hat U_{\tar, k} = \mathcal{T}\exp\{-i \int_{k\delta T}^{(k+1)\delta T}\hat H_\dr[\alpha(t)] \dif t\}$. Recall that $\tilde \alpha(t) = \alpha(t) f(t)$, where $f(t)$ is a periodic function with period $\delta T$, scales as $O(1)$, and satisfies Eq.~(\ref{f_even}, \ref{f_avg}). Without loss of generality, we can just focus on $k=0$ case and calculate $\hat U_0 \hat U_{\tar, 0}^\da$, since $\hat U_k \hat U_{\tar, k}^\da$ with any $k$ has a similar structure as the situation with $k=0$. By analogy with the Magnus expansion, we have 
\begin{widetext}
\begin{equation}\label{BCHexponential}
\begin{split}
    \hat U_0 \hat U_{\tar, 0}^\da = \exp\bigg( & -i\int_0^{\delta T} \dif t\ [\hat H'_\dr(t) - \hat H_\dr(t)]  - \frac{1}{2}\int_0^{\delta T} \dif t_2 \int_0^{t_2} \dif t_1 \{[\hat H'_\dr(t_2), \hat H'_\dr(t_1)] - [\hat H_\dr(t_2), \hat H_\dr(t_1)]\} \\
    & + \frac{1}{2} \int_0^{\delta T} \dif t_2\int_0^{\delta T} \dif t_1 [\hat H'_\dr(t_2), \hat H_\dr(t_1)] + \hat R_3 \bigg).
\end{split}
\end{equation}
\end{widetext}
It is easy to see that $\hat R_3$ is anti-Hermitian and $\hat R_3 \sim O[(\chi\alpha^2 \delta T)^3]$. As we discussed in the main text, if $\alpha(t)$ is a constant within $t \in [0, \delta T]$, the first and the second order terms written explicitly in Eq.~\eqref{BCHexponential} are zero. But here we want to talk about the more general case that $\alpha(t)$ is time dependent. Even in this situation, we will show that the contribution of the error from the first two orders are small compared with the third order term.

We can write $\alpha(t)$ as
\begin{equation}
\begin{split}
    \alpha(t) ={}& \alpha(t_0) + \dot \alpha(t_0)(t - t_0) + \frac{1}{2} \ddot\alpha(t_0)(t - t_0)^2\\
    & + O[(t - t_0)^3],
\end{split}
\end{equation}
where $t_0 = \frac{\delta T}{2}$. Since we assumed that $\alpha(t)$ is a slowly varied function, we have $\dot \alpha \sim O(\alpha/T)$ and $\ddot \alpha \sim O(\alpha/T^2)$. As a result, based on the structure of $f(t)$, for the first order term we have
\begin{equation}
    \begin{split}
        &\int_0^{\delta T} \dif t\ \chi[\tilde\alpha(t) - \alpha(t)] \sim O\left[\frac{\chi\alpha (\delta T)^3}{T^2}\right],\\
        &\int_0^{\delta T} \dif t\ \chi\tilde\alpha^2(t) \sim O\left[\frac{\chi\alpha^2 (\delta T)^3}{T^2}\right].
    \end{split}
\end{equation}
Since we have assumed that $T \sim O(1/\chi\alpha)$, both of the two terms are smaller compared with $O[(\chi\alpha^2\delta T)^3]$ when $\alpha$ is large.

A similar analysis can be performed for the second-order terms. The dominant part lies in $[\hat H'_\dr(t_2), \hat H'_\dr(t_1)]$ since it contains $O[(\chi\alpha^2)^2]$ coefficients. We find that
\begin{equation}
\begin{split}
    &\int_0^{\delta T} \dif t_2 \int_0^{t_2} \dif t_1 \ \chi^2[\tilde \alpha^2(t_2) \tilde\alpha^{*2}(t_1) - \tilde \alpha^{*2}(t_2) \tilde\alpha^{2}(t_1)]\\
    &\sim O\left(\frac{\chi^2\alpha^4(\delta T)^3}{T}\right),
\end{split}
\end{equation}
which is also a higher order term compared with $O[(\chi\alpha^2 \delta T)^3]$. So, the dominant scaling of Trotter error does not change even if $\alpha(t)$ is a slowly varied function introduced in Eq.~\eqref{alpha_t}.

\section{Errors from inaccurate control}\label{app:inaccu_ctrl}

In this part, we will investigate the extra infidelity induced by inaccurate control of $\Lambda_1(t)$. As shown in Sec.~\ref{app:err_ana}, $|\Lambda_1| \sim O(\alpha M/T)$ can be strong enough so that a deviation for a small portion of it may cause a huge impact. Here we consider the problem that how the infidelity changes if the actual driving pulse we use is $(1+\eta) \Lambda_1(t)$ where $\eta$ is a small dimensionless number.

We first briefly derive the contribution of the extra $\eta \Lambda_1(t)$ to the dynamics. Notice that, if we go to the same displaced rotating frame as mentioned in the main text, now the Hamiltonian should be written as 
\begin{equation}\label{eqn:hamil_eta}
\hat H_\eta = \hat H'_\dr[\tilde{\alpha}(t)] + [\eta \Lambda_1(t) \hat a^\da + \hc],
\end{equation}
where $\hat H'_\dr[\tilde{\alpha}(t)]$ has been defined in Eq.~\eqref{eqn:new_single_drive} and $\Lambda_1(t) = \chi\tilde\alpha(t)[|\tilde\alpha(t)|^2 - r] + i\kappa\tilde{\alpha}(t)/2 + i\dot {\tilde\alpha}(t)$ (as in Eq.~\eqref{eq:Lambda1_tilde}). We can again perform a frame transformation such that $\hat a \to \hat a + \beta(t)$ where $\beta(t)$ satisfies 
\begin{equation}
\dot \beta(t) + \frac{\kappa}{2} \beta(t) = -i\eta \Lambda_1(t).
\end{equation}
Further, due to the specific form of $\Lambda_1(t)$, we can write $\beta(t) = \beta_1(t) + \beta_2(t)$ where $\beta_1(t) = \eta \tilde{\alpha}(t)$ and $\beta_2(t) = -i\eta \int_0^t e^{-\kappa(t-t')/2} \chi \tilde\alpha(t')[|\tilde\alpha(t')|^2 - r] \dif t'$. It is worth mentioning that, although $i\dot{\tilde{\alpha}}$ dominates in $\Lambda_1$, its contribution to $\beta$ is much smaller compared with the contribution from the $O(\chi \alpha^3)$ term, provided that $\alpha \gg 1$.

Now the Hamiltonian in the new frame should be
\begin{equation}\label{eqn:hamil_eta_pr}
\begin{split}
\hat H'_\eta ={} & \frac{\chi}{2} (\hat a^\da + \beta^*)^2 (\hat a + \beta)^2\\
& + \{\chi \tilde{\alpha}(\hat a^\da + \beta^*)[(\hat a^\da + \beta^*)(\hat a + \beta) - r] + \hc\}\\
& + [\frac{\chi}{2} \tilde \alpha^2 (\hat a^\da + \beta^*)^2 + \hc].
\end{split}
\end{equation}
Clearly $\hat H'_\eta$ deviates from the ideal $\hat H'_\dr[\tilde \alpha(t)]$ as $\beta$ increases from zero. Besides, in general $\beta(T) \neq 0$, which suggests that we need an extra displacement operation $\hat D[\beta(T)]$ to let the states go back to the non-displaced frame. But in practice we do not know what the $\beta(T)$ is and will not actively apply this operation, which may lead to extra errors.

\begin{figure}[t!]
\centering
\includegraphics[scale = 0.49]{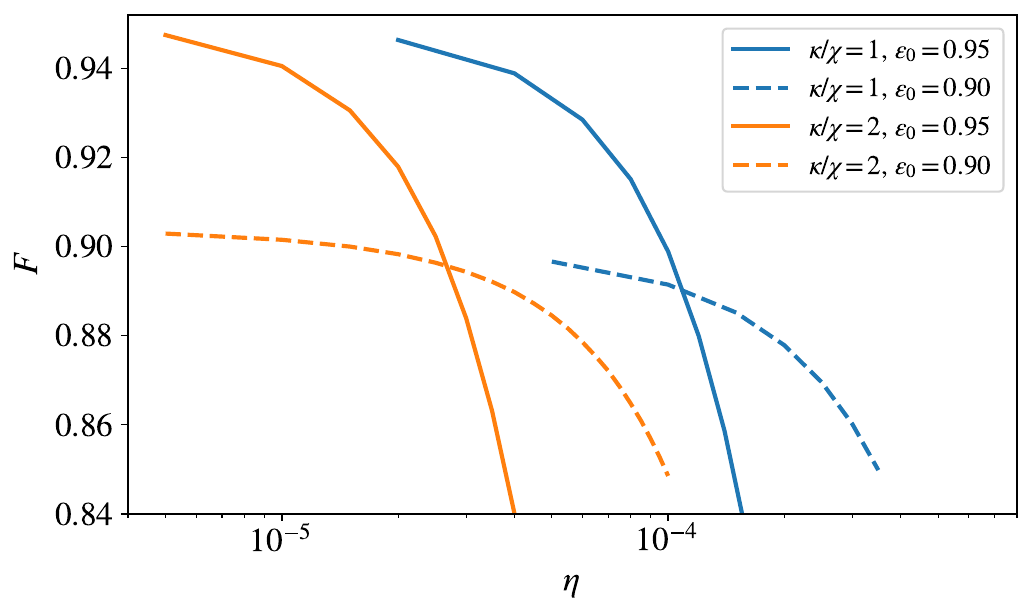}
\caption{Fidelity changes due to the amplitude deviation with the ideal pulses for the Fock $\ket{1}$ state preparation task, as explained in Sec.~\ref{app:inaccu_ctrl}.}
\label{fig:dr_dev}
\end{figure}

Finally we try to numerically consider a specific example, which is the Fock $\ket{1}$ state preparation with constant $\alpha$ as mentioned in Sec.~\ref{app:err_ana}. We assume that, $T$ and $M$ are chosen optimally such that for a given target infidelity $\epsilon_0$ we have $c_1 \kappa T = \frac{14}{15}\epsilon_0$ and $\frac{c_2}{M^4(\chi T)^6} = \frac{1}{15} \epsilon_0$. However, during the constant $\alpha$ evolution time the actual 1-photon driving amplitude we implement is $(1+\eta)\Lambda_1(t)$. From FIG.~\ref{fig:dr_dev} we can see how the fidelity $F = 1 - \epsilon$ decreases as the increase of the inaccuracy $\eta$. We notice that for a smaller $\epsilon_0$ expected, we need a larger $\alpha$. However, larger $\alpha$ will result in larger deviation $\beta$. The fidelity will be more sensitive to $\eta$ with larger $\alpha$ and therefore it decays faster as $\eta$ grows when $\epsilon_0$ is small.

\section{Dynamics beyond rotating wave approximation}\label{app:self_con}

\subsection{Overview}

In the main text, we did not talk about any property related to the resonator frequency $\omega_c$, since it simply disappears after we go to the rotating frame and perform the rotating wave approximation (RWA). In practice, however, the validity of RWA will set a limitation on our current protocol. If we can do either charge or flux drive but not both, then the driving amplitude $\Lambda_1$ cannot be too strong, which further limits the maximum $|\alpha|$ and the oscillation frequency $\omega_r$ for $\at(t)$ that is used to suppress the infidelity. In this section, we will discuss the effects of those off-resonant terms in the aspect of scaling, and we also provide some strategies to partially compensate for them. 

Before we start to talk about the specific problem, we will briefly explain the analysis strategy we are going to use. Rather than going to the displaced rotating frame we mentioned in the main text, which is treated as the 0-th order reference frame here, we try to find another reference frame with higher order corrections on the original one and compare the difference between Hamiltonian there and the one ($\hat H'_\dr[\tilde{\alpha}(t)]$) we desire to engineer. To make sure the dynamics induced by the difference between the two Hamiltonian are negligible, we need higher-order corrections on the 1-photon driving amplitude $\Lambda_1(t)$ and frequency $\omega_1(t)$, together with several assumptions on the relationship among some relevant parameters (see Eq.~\eqref{eqn:assum} for example).

As we shall see, there are two kinds of terms in the difference between desired and actual Hamiltonian. One kind includes terms that have a fast-oscillating coefficient, which is usually proportional to $e^{ik\omega_c t}$ or $e^{ik\omega_r t}$ with nonzero integer $k$; while another kind contains those terms without fast-oscillating factors, which are called as ``slowly varied" terms. Besides, we also call terms with coefficients $e^{ik\omega_r t}$  ``slowly varied" when they are compared with $e^{ik\omega_c t}$ (or we call ``slowly varied" compared with $\omega_c$ for short).

For both cases, the dynamics induced by those terms is negligible if the amplitudes (absolute value) of their coefficients are too small such that the integration of the amplitudes over gate time $T$ is far less than $1$. For example, terms with slowly varied coefficient $c(t)$ or the fast oscillating $c(t)e^{i\omega t}$ can be both ignored if $\int_0^T |c(t)| \dif t \ll 1$. However, for terms with fast oscillating coefficients that do not belong to the former situation, we need to consider some effective Hamiltonian corresponding to those terms. For example, the dynamics generated by a Hamiltonian like $\hat H = \hat A e^{i\omega t} + \hat A^\da e^{-i\omega t}$ can be approximated by that generated via an effective Hamiltonian $\hat H_\text{eff} = [\hat A, \hat A^\da]/\omega$, provided that $\omega$ is much larger than some specific matrix norm of $\hat A$. This can be simply demonstrated via Magnus expansion. In this way, we will focus on the scaling of the coefficients for those slowly varied effective counterparts and see if the time integration of those is far less than $1$ and therefore negligible as well. Similarly, remember that in the parameter regime we want to work in, the coherent error induced by $(\frac{\chi}{2} \at^2 \hat a^{\da 2} + \hc)$ is negligible. In general, for terms that oscillate in the form of $f^2(t)$ (where $f(t)$ is a periodic function with frequency $\omega_r$ and satisfies Eq.~(\ref{f_even}, \ref{f_avg})), their contribution to the dynamics can be ignored even when their amplitudes scale as $O(\chi \alpha^2)$. We will also make use of this fact later. These discussions provide support for the following derivations.

Our scaling analysis also relies on some assumptions for the ideal blockade Hamiltonian. We assume the dimension of the blockade subspace is small, such that in the scaling analysis the matrix elements of $\hat a$ and $\hat a^\da$ are of the order $O(1)$ when dynamics is restricted into the blockade subspace. Also, we assume that the control function $\alpha(t)$ is slowly varied in time, which (as mentioned in Sec.~\ref{AppTimeDep}) implies that the time derivative of $\alpha(t)$ scales as $\dot\alpha(t) \sim O(\alpha/T)$, and similarly $\dif{}^{n} \alpha(t) / \dif t^n \sim O(\alpha/T^n)$.

\subsection{Detailed scaling analysis on the dynamics}

In general, the Hamiltonian in the lab frame without RWA should be written as
\begin{equation}\label{eqn:H_nr_lab}
\begin{split}
\hat H ={} & \frac{\chi}{12} (\hat a^{\da}+\hat a)^4 + (\omega_c - \chi) \hat a^\da \hat a\\
& + [\Lambda_1(t) e^{-i\omega_1(t) t} + \Lambda^*_1(t) e^{i\omega_1(t) t}](\hat a^\da + \hat a).
\end{split}
\end{equation}
From then on we denote $\phi_d(t) := \omega_1(t)t - \omega_c t$ and as shown in the main text we have $\dot \phi_d(t) \sim O(\chi \alpha^2)$. If we still go to the displaced rotating frame shown in the main text, the Hamiltonian will become
\begin{widetext}
\begin{equation}\label{eqn:H_nr}
\begin{split}
\hat H_\NR ={}& \frac{\chi}{2} (\hat a^\da + \at^*)^2 (\hat a + \at)^2 - \dot \phi_d(t) (\hat a^\da + \at^*)(\hat a + \at) + [(\Lambda_1(t) - i\dot\at - i\kappa\at/2) \hat a^\da + \hc]\\
& + (\Lambda_1^*(t) e^{2i\phi_d(t)}e^{2i\omega_c t} \hat a^\da + \hc)\\
& + \left\{\left[\frac{\chi}{3}  (\hat a^\da + \at^*)^3 (\hat a + \at) + \frac{\chi}{2}  (\hat a^\da + \at^*)^2\right] e^{2i\phi_d(t)}e^{2i\omega_c t} + \frac{\chi}{12}  (\hat a^\da + \at^*)^4 e^{4i\phi_d(t)}e^{4i\omega_c t} + \hc\right\}.
\end{split}
\end{equation}
\end{widetext}
It can be easily seen that we can recover the Hamiltonian under RWA by throwing away all the fast oscillating terms related to $e^{i\omega_c t}$. Therefore, as proposed earlier we can choose $\Lambda_1^\zth$ and $\phi^\zth_d$ to achieve the desired Hamiltonian under RWA by
\begin{equation}
\begin{split}
\Lambda_1^\zth &= \dot \phi^\zth_d \at^\zth - \chi \at^\zth (|\at^\zth|^2 + r) + i \dot \at^\zth + i \kappa\at^\zth/2, \\
\dot \phi^\zth_d &= 2\chi |\at^\zth|^2.
\end{split}
\end{equation}
where $\at^\zth$ is the $\at$ function we designed in the main text.

Now we want to focus on those fast oscillating terms. One way is to go to a slightly different displaced frame to absorb some large components. Note that $\Lambda_1 \sim i \dot \at \sim \alpha \omega_r \gg \chi \alpha^3$, which means $(\Lambda_1^*(t) e^{2i\phi_d(t)}e^{2i\omega_c t} \hat a^\da + \hc)$ has the largest amplitude and we want to deal with it first. Basically, we want to choose $\at = \at^\zth + \at^\fst$ such that the amplitude of the linear $\hat a^\da$ term can be mostly absorbed in $\at^\fst$. We first introduce $\at^\fpst$ tentatively such that
\begin{widetext}
\begin{equation}\label{eqn:at_fp_def}
\begin{split}
i\dot \at^\fpst + i\kappa \at^\fpst/2 ={}& [\Lambda_1^{*[0]} + \chi \at^{*[0]} (|\at^\zth|^2 + 1)] e^{2i\phi^\zth_d(t)}e^{2i\omega_c t}\\
 &+ \frac{\chi}{3} (\at^{*[0]})^3 e^{4i\phi^\zth_d(t)}e^{4i\omega_c t} + \frac{\chi}{3} (\at^\zth)^{3} e^{-2i\phi^\zth_d(t)}e^{-2i\omega_c t}.
\end{split}
\end{equation}
\end{widetext}
For simplicity, we denote the right-hand-side (RHS) of the equation above as $g(t)$.


The solution for $\at^\fpst$ can be written down explicitly as
\begin{equation}
\at^\fpst(t) = -i\int_0^t e^{-\kappa (t - t')/2} g(t') \dif t'.
\end{equation}
Via the technique of integration by parts, we have the following identity
\begin{equation}\label{eqn:int_by_part}
\begin{split}
 \int_0^t h(t') e^{i\omega t'} \dif t' ={}& \sum_{n=0}^k \left( \frac{i}{\omega} \right)^{n+1} [h^{(n)}(0) - h^{(n)}(t) e^{i\omega t}] \\
& + \left( \frac{i}{\omega} \right)^{k+1} \int_0^t h^{(k+1)}(t') e^{i\omega t'} \dif t',
\end{split}
\end{equation}
where $h^{(n)}(t) = \frac{\dif{}^n}{\dif {t}{}^n} h(t)$. In Eq.~\eqref{eqn:int_by_part} we have implicitly assumed that $h^{(n)}(t)$ is a continuous function for $n \leq k$. Notice that if for any $m$, $\frac{1}{\omega} |\frac{\dif}{\dif t} \ln h^{(m)}(t)| = \frac{1}{\omega}\frac{|h^{(m+1)}(t)|}{|h^{(m)}(t)|} \ll 1$, then the summation in the first line of Eq.~\eqref{eqn:int_by_part} can be convergent quickly and the residual terms can be ignored. In our situation here when calculating $\at^\fpst(t)$, we need to assume $\omega_r/\omega_c \ll 1$ since $\at^\zth (t)$ and $\Lambda_1^\zth (t)$ contains frequency components related to $\omega_r$. In fact, as mentioned in former sections we always assume that $\kappa \ll \chi \alpha \ll \dot\phi_d \sim \chi \alpha^2 \ll \omega_r$ to guarantee high-fidelity operations, which means the assumption of $\omega_r/\omega_c \ll 1$ implies $\kappa/\omega_c \ll 1$ and $\chi \alpha^2 /\omega_c \ll 1$. Later on, we simply ignore contributions from the factor $e^{-\kappa(t-t')/2}$ since we want to work in the regime that $\kappa t \leq \kappa T \ll 1$ to achieve high fidelity. Also, it varies slowly over time, so it will not affect any scaling analysis when taking time derivatives of this.

With the trick mentioned above, we can write $\at^\fpst$ in the following form:
\begin{equation}
\at^\fpst = \at^{[1',0]} + \at^{[1',1]} e^{2i\omega_c t} + \at^{[1',2]} e^{4i\omega_c t} + \at^{[1',-1]} e^{-2i\omega_c t}.
\end{equation}
We first analyze $\at^{[1',1]}$ since $\Lambda_1^{*[0]}$ dominates in $g(t)$. For the lowest order term ($n = 0$) in Eq.~\eqref{eqn:int_by_part}, notice that $\Lambda_1^\zth$ contains $i\dot\at^\zth$ term where $\tilde{\alpha}^\zth(t) = \alpha(t) f(t)$, we have the leading contribution in $\at^{[1',1]}$ as $\at^{[1',1]} \simeq \frac{i\alpha^* \dot f^*}{2\omega_c} e^{2i\phi_d^\zth} \sim O(\frac{\alpha \omega_r}{\omega_c})$. It also contains $O(\frac{\chi \alpha^3}{\omega_c})$ terms that come from those $O(\chi \alpha^3)$ parts in $g(t)$ and those implicitly in $\Lambda_1^\zth$. Also, since we assumed $\alpha(t)$ varies slowly that $\dot \alpha(t) \sim O(\alpha / T) \sim O(\chi \alpha^2)$, it gives a correction to the estimation of $\at^{[1',1]}$ with $O(\frac{\dot \alpha f}{\omega_c}) \sim O(\frac{\chi \alpha^2}{\omega_c})$. Besides, the second order term ($n = 1$) in Eq.~\eqref{eqn:int_by_part} can give a contribution of $O(\frac{\alpha \ddot f}{\omega_c^2}) \sim O(\frac{\alpha \omega_r^2}{\omega_c^2})$. Using similar argument, we can show that $\at^{[1',2]} \sim O(\frac{\chi \alpha^3}{\omega_c})$ and $\at^{[1',-1]} \sim O(\frac{\chi \alpha^3}{\omega_c})$. The analysis for $\at^{[1',0]}$ is a little different. As mentioned in the main text, in practice we start with $\alpha(0) = 0$ while $\dot \alpha \sim O(\alpha/T) \sim O(\chi \alpha^2)$. Therefore, we have $\Lambda_1^\zth(0) \sim O(\chi \alpha^2)$ so that $\at^{[1',0]} \sim O(\frac{\chi \alpha^2}{\omega_c})$. Other terms from $g(t)$ will only cause higher order corrections.


Later we want to show how we can partially compensate for the effects of those fast-oscillating terms and even with this technique we still need further assumptions that
\begin{equation}\label{eqn:assum}
\frac{\chi \alpha^3}{\omega_c} \ll 1, \quad \frac{\alpha \omega_r^2}{\omega_c^2} \ll 1,
\end{equation}
so that the residual effects could be small.

We focus on the change of the Hamiltonian under the new displaced frame with $\at = \at^\zth + \at^\fpst$. We want first to check those extra terms (in comparison with $\hat H'_\dr[\tilde{\alpha}(t)]$) which are induced by $\at^\fpst$ and do not contain fast oscillating factor (factor $e^{ik\omega_c t}$ where $|k| \in \mathbb{N}_+$). We try to show the dominant effect from each term in Eq.~\eqref{eqn:H_nr}, and evaluate their contributions under assumptions Eq.~\eqref{eqn:assum}.

For the corrections with cubic $\hat a$ or $\hat a^\da$ terms (including $\hat a^{\da 3}$, $\hat a^{\da 2}\hat a$ and their Hermitian conjugate), their coefficients will be of $O(\chi \at^\fpst)$, whose amplitudes scale at most $O(\chi \alpha \frac{\omega_r}{\omega_c})$. Even without the fast-oscillating factor, the integration of their amplitudes over time will give us $\chi \alpha T \frac{\omega_r}{\omega_c} \sim \frac{\omega_r}{\omega_c} \ll 1$ (where we have used $T \sim \frac{1}{\chi\alpha}$), which means the contribution from those correction terms to the dynamics are small.

For the correction induced by slowly varied (in comparison with $\omega_c$) $\hat a^\da \hat a$ terms, it can come from either $2\chi |\at|^2 \hat a^\da \hat a$ or $(\chi \at^{*2} e^{2i\phi^\zth_d} e^{2i\omega_c t} + \cc) \hat a^\da \hat a$, where $\cc$ stands for ``complex conjugate". The former will lead extra slowly varied $\hat a^\da \hat a$ terms with coefficients like $(2\chi \at^\zth \at^{*[1',0]} + \cc) \sim O(\frac{1}{T} \frac{\chi \alpha^2}{\omega_c})$ and $2\chi |\at^{[1',1]}|^2 \sim O(\frac{1}{T} \frac{\alpha\omega_r^2}{\omega_c^2})$, while the coefficient of the dominant correction from the later will be $(2\chi \at^{*[0]} \at^{*[1',1]} e^{2i\phi^\zth_d}+ \cc) \sim O(\chi \alpha^2 \frac{\omega_r}{\omega_c})$. We introduce $\phi^\fst_d(t)$ function and make its time derivative $\dot\phi^\fst_d(t)$ to be equal to the summation of all the slowly varied coefficients of the correction terms, as
\begin{widetext}
\begin{equation}\label{eqn:dot_phi_d1_def}
\dot\phi^\fst_d(t) = 2\chi \left\{\sum_{k=-1}^2 |\at^{[1',k]}|^2 + (\at^\zth \at^{*[1',0]} + \cc) + \{[(\at^{*[0]}+\at^{*[1',0]}) \at^{*[1',1]} + \at^{*[1',2]} \at^{*[1',-1]}] e^{2i\phi^\zth_d} + \cc\}\right\}.
\end{equation}
\end{widetext}
Apparently we still have $\dot \phi^\fst_d \sim O(\chi \alpha^2 \frac{\omega_r}{\omega_c})$. This term may not be ignored and later we will show that we can compensate for that by changing the frequency $\omega_1(t)$ of the linear drive.

Then we discuss the corrections for slowly varied (in comparison with $\omega_c$) $\hat a^{\da 2}$ terms which come from $\frac{\chi}{2}(\at + \at^*e^{2i\phi^\zth_d} e^{2i\omega_c t})^2 \hat a^{\da 2}$. Those corrections are
\begin{widetext}
\begin{equation}\label{eqn:ad2_corr}
\begin{split}
    \frac{\chi}{2} \{&(2\at^\zth + \at^{[1',0]} + \at^{*[1',1]}e^{2i\phi^\zth_d})(\at^{[1',0]} + \at^{*[1',1]}e^{2i\phi^\zth_d})\\
    &+ 2[\at^{*[1',1]} + (\at^{*[0]} + \at^{*[1',0]})e^{2i\phi^\zth_d}](\at^{*[1',-1]} + \at^{*[1',2]} e^{2i\phi^\zth_d})\} \hat a^{\da 2}.
\end{split}
\end{equation}
\end{widetext}
The dominant coefficient of the correction will be $\chi\at^\zth \at^{*[1',1]}e^{2i\phi^\zth_d} \sim O(\frac{\chi \alpha^2 \omega_r}{\omega_c})$. Although the amplitude of this term might be large, we notice that $\at^{[1',1]} \simeq \frac{i\alpha^* \dot f^*}{2\omega_c} e^{2i\phi_d^\zth}$ with corrections of the order $O(\frac{\chi\alpha^2}{\omega_c})$ and $O(\frac{\alpha\omega_r^2}{\omega_c^2})$. As a result, $\chi\at^\zth \at^{*[1',1]}e^{2i\phi^\zth_d} \simeq -i\frac{\chi \alpha^2 f \dot f}{2\omega_c}$. We notice that $\frac{f\dot f}{\omega_c}$ is a periodic function with period $\delta T = \frac{2\pi}{\omega_r}$. By averaging within each period, we have $\overline{f\dot f} = \int_0^{\delta T} f\dot f \dif t = 0$. So, $f\dot f$ is a linear superposition of fast oscillating $e^{ik\omega_r t}$ functions with nonzero integer $k$. We can now use the effective Hamiltonian argument we brought up previously. The commutator between this and the 0-th order $[\chi (\at^\zth)^2 \hat a^{\da 2} + \hc]$ will give an effective term with amplitude scaled as $\frac{1}{\omega_r}(\frac{\chi \alpha^2 \omega_r}{\omega_c})(\chi \alpha^2) \sim \frac{\chi \alpha^3}{\omega_c}\frac{1}{T}$, which is negligible after integration over time $T$ under our assumptions Eq.~\eqref{eqn:assum}. However, this trick does not apply to higher order correction terms of $\at^{*[1',1]}$ like $-\frac{\alpha \ddot f}{4\omega_c^2} \sim O(\frac{\alpha\omega^2_r}{\omega^2_c})$, since it may lead to slowly varied corrections (compared with $\omega_r$) in $\chi\at^\zth \at^{*[1',1]}e^{2i\phi^\zth_d}$. But with the second assumption in Eq.~\eqref{eqn:assum}, after integration over time we still have $\chi \alpha \frac{\alpha\omega^2_r}{\omega^2_c} T \ll 1$, which means the contribution from the higher order correction of $\at^{*[1',1]}$ can still be ignored. We can similarly argue that other terms in Eq.~\eqref{eqn:ad2_corr} can be ignored since the amplitudes of their coefficients are also too small.

Now we start the procedure of partial compensation. We adjust the 1-photon driving frequency such that $\phi_d = \phi_d^\zth + \phi_d^\fst$. Therefore, we choose a new correction $\at^\fst$ for the displaced frame which satisfies
\begin{widetext}
\begin{equation}\label{eqn:at_f_def}
\begin{split}
i\dot \at^\fst + i\kappa \at^\fst/2 ={}& [\Lambda_1^{*[0]} + \chi \at^{*[0]} (|\at^\zth|^2 + 1)] e^{2i[\phi^\zth_d(t) + \phi^\fst_d(t)]}e^{2i\omega_c t}\\
 &+ \frac{\chi}{3} (\at^{*[0]})^3 e^{4i[\phi^\zth_d(t) + \phi^\fst_d(t)]}e^{4i\omega_c t} + \frac{\chi}{3} (\at^\zth)^{3} e^{-2i[\phi^\zth_d(t) + \phi^\fst_d(t)]}e^{-2i\omega_c t}.
\end{split}
\end{equation}
\end{widetext}
The only difference between here and Eq.~\eqref{eqn:at_fp_def} is that we use $e^{2ki[\phi^\zth_d(t) + \phi^\fst_d(t)]}$ instead of $e^{2ki\phi^\zth_d(t)}$. We denote
\begin{equation}
\at^\fst = \at^{[1,0]} + \at^{[1,1]} e^{2i\omega_c t} + \at^{[1,2]} e^{4i\omega_c t} + \at^{[1,-1]} e^{-2i\omega_c t}.
\end{equation}
Then, from Eq.~\eqref{eqn:int_by_part} we can find that
\begin{equation}
\begin{split}
\at^{[1,1]} &= \at^{[1',1]} e^{2i\phi_d^\fst(t)} + O(\frac{\Lambda_1}{\omega_c}\frac{\dot \phi_d^\fst}{\omega_c})\\
&= \at^{[1',1]} e^{2i\phi_d^\fst(t)} + O(\frac{\chi \alpha^3 \omega_r^2}{\omega^3_c}).
\end{split}
\end{equation}
Similarly, we can also find that the difference between $\at^{[1,k]}$ and $\at^{[1',k]} e^{2ik\phi_d^\fst(t)}$ (for $k=-1, 0, 2$) are even smaller compared with $O(\frac{\chi \alpha^3 \omega_r^2}{\omega^3_c})$ in the scaling aspect. The difference between $\at^\fst$ and $\at^\fpst$ can lead to extra non-fast-oscillating (compared with $\omega_c$) terms. For example, the extra coefficient induced in this way for $\hat a^\da \hat a$ will be at most the same scaling as $[2\chi \at^{*[0]}(\at^{*[1,1]}e^{2i(\phi_d^\zth + \phi_d^\fst)} - \at^{*[1',1]}e^{2i\phi_d^\zth}) + \cc] \sim O(\frac{\chi \alpha^3 \omega_r^2}{T\omega^3_c})$, which will be far less that $1$ and therefore can be ignored after integration over time. We can use the same way to argue the extra effect from non-fast-oscillating correction of $\hat a^{\da 2}$ term can be ignored, due to the similarity in the structure of $\at^\fst$ and $\at^\fpst$.

The coefficient of slowly varied correction (in comparison with $\omega_c$) for $\hat a^\da$ term will be at most $O(\chi \alpha^2 \at^{[1,1]}) \sim O(\chi \alpha^3 \frac{\omega_r}{\omega_c})$. To deal with it, we introduce a correction for $\Lambda_1 = \Lambda_1^\zth + \Lambda_1^\fst$, where
\begin{widetext}
\begin{equation}
\begin{split}
\Lambda_1^\fst ={}& -\chi [(|\at^\zth + \at^{[1,0]}|^2 + |\at^{[1,1]}|^2 + |\at^{[1,2]}|^2 + |\at^{[1,-1]}|^2)(\at^\zth + \at^{[1,0]} + \at^{*[1,1]}e^{2i\phi_d}) - |\at^\zth|^2 \at^\zth]\\
&-\chi[\at^{[1,1]}(\at^{*[0]} + \at^{*[1,0]}) + (\at^\zth + \at^{[1,0]})\at^{*[1,-1]} + \at^{[1,2]}\at^{*[1,1]}](\at^{[1,-1]}+\at^{*[1,2]}e^{2i\phi_d})\\
&-\chi[\at^{*[1,1]}(\at^{[0]} + \at^{[1,0]}) + (\at^{*[0]} + \at^{*[1,0]})\at^{[1,-1]} + \at^{*[1,2]}\at^{[1,1]}][\at^{[1,1]}+(\at^{*[0]} + \at^{*[1,0]})e^{2i\phi_d}]\\
&-\chi[(\at^{[0]} + \at^{[1,0]})\at^{*[1,2]} + \at^{[1,-1]}\at^{*[1,1]}](\at^{[1,2]}+\at^{*[1,-1]}e^{2i\phi_d}) + \chi \at^{*[1,1]}e^{2i\phi_d}\\
&+[\dot\phi_d^\zth \at^{[1,0]} + \dot\phi_d^\fst(\at^\zth + \at^{[1,0]})]\\
&-\chi [\at^{*[1,2]}(\at^{*[0]} + \at^{*[1,0]})^2 + 2\at^{*[1,2]}\at^{*[1,1]}\at^{*[1,-1]} + (\at^{*[1,1]})^2 (\at^{*[0]} + \at^{*[1,0]})]e^{4i\phi_d} \\
&-\chi [2\at^{[1,-1]}(\at^{[0]} + \at^{[1,0]})\at^{[1,2]} + \at^{[1,-1]}(\at^{[1,1]})^2 + (\at^{[0]} + \at^{[1,0]})^2\at^{[1,1]}]e^{-2i\phi_d}.
\end{split}
\end{equation}
\end{widetext}
Here $\Lambda_1^\fst \sim O(\chi \alpha^3 \frac{\omega_r}{\omega_c})$ is chosen  to fully absorb those non-fast-oscillating corrections. It is still worth to mention that after this change there will be extra $\Lambda^{*[1]}_1 e^{2i\phi_d(t)}e^{2i\omega_c t}$ and some other fast-oscillating terms with amplitude scaled at most $O(\chi \alpha^3 \frac{\omega_r}{\omega_c})$ in the new Hamiltonian, since all others with larger amplitudes have been absorbed due to Eq.~\eqref{eqn:at_f_def}.

Further, we can introduce a correction $\at^{[2]} \sim O(\chi \alpha^3 \frac{\omega_r}{\omega^2_c}) \ll 1$ and repeat all the steps we did (including $\phi_d^{[2]}$ and $\Lambda_1^{[2]}$ compensation) so that the contribution from $\at^{[2]}$ are too small and we can ignore this. After $k$ rounds of all those corrections, terms $\hat a^\da$ with coefficients oscillating in the speed of $\omega_c$ will be at most $O[\alpha\omega_r(\frac{\chi \alpha^2}{\omega_c})^k]$, which will be small compared with $\frac{1}{T}$ for $k \geq 3$ under the assumptions in Eq.~\eqref{eqn:assum}.

Finally we talk about the effects from those fast-oscillating terms (under $\omega_c$) of $\hat a^{\da 2}$, $\hat a^\da \hat a$ and $\hat a^2$ in Eq.~\eqref{eqn:H_nr}, whose coefficients are at most $O(\chi \alpha^2)$. We denote the summation of them as $\hat H_{\NR, 2}(t)$, which can be written as 
\begin{widetext}
\begin{equation}
\begin{split}
\hat H_{\NR, 2}(t) ={}& \Big\{\Big[\frac{\chi}{2} [(2|\at^\zth|^2 + 1) e^{2i\phi_d} e^{2i\omega_c t} + (\at^{*[0]})^2 e^{4i\phi_d} e^{4i \omega_c t}] + O(\frac{\chi \alpha^2 \omega_r}{\omega_c} e^{2ik\omega_c t})\Big] \hat a^{\da 2} + \hc\Big\} \\
&+ \Big[[\chi (\at^{*[0]})^2 e^{2i\phi_d} e^{2i\omega_c t} + \cc] + O(\frac{\chi \alpha^2 \omega_r}{\omega_c} e^{2ik\omega_c t}) \Big] \hat a^\da \hat a .
\end{split}
\end{equation}
\end{widetext}

We denote $\hat U_{\NR, 2}(t) := \mathcal{T} \exp(-i\int_0^t \hat H_{\NR, 2}(t') \dif t')$, then for any Hamiltonian $\hat H$ we can perform a frame transformation to get a new one $\hat H' = \hat U^\da_{\NR, 2} \hat H \hat U_{\NR, 2} + i \dot{\hat U}^\da_{\NR, 2} \hat U_{\NR, 2} = \hat U^\da_{\NR, 2} (\hat H - \hat H_{\NR, 2}) \hat U_{\NR, 2}$. We can choose $\hat H$ as the Hamiltonian after going to the displaced rotating frame with all rounds of corrections. To achieve $\hat H'$, we can replace $\hat a$ with $\hat a(t) := \hat U^\da_{\NR, 2}(t) \hat a \hat U_{\NR, 2}(t)$ and $\hat a^\da$ with $\hat a^\da(t) := \hat U^\da_{\NR, 2}(t) \hat a^\da \hat U_{\NR, 2}(t)$ in $(\hat H - \hat H_{\NR, 2})$ (also if we consider loss we need to change $\mathcal{D}[\hat a]$ into $\mathcal{D}[\hat a(t)]$). From the definition of $\hat a(t)$ and $\hat a^\da(t)$, we have
\begin{widetext}
\begin{equation}\label{eqn:Heisenberg_a}
\begin{split}
\frac{\dif \hat a(t)}{\dif t} ={}& (-i)[\chi (2|\at^\zth|^2 + 1) e^{2i\phi_d} e^{2i\omega_c t} + \chi (\at^{*[0]})^2 e^{4i\phi_d} e^{4i\omega_c t} + O(\frac{\chi \alpha^2 \omega_r}{\omega_c} e^{2ik\omega_c t})]\hat a^\da(t)\\
&- i\Big[[\chi (\at^{*[0]})^2 e^{2i\phi_d} e^{2i\omega_c t} + \cc] + O(\frac{\chi \alpha^2 \omega_r}{\omega_c} e^{2ik\omega_c t}) \Big] \hat a(t),
\end{split}
\end{equation}
\end{widetext}
and also $\dot{\hat a}^\da(t) = [\dot{\hat a}(t)]^\da$ is also equal to a linear superposition of $\hat a^\da(t)$ and $\hat a(t)$. From scaling analysis we know that $\frac{(\chi \alpha^2)^2}{\omega_c}T \sim \frac{\chi \alpha^3}{\omega_c} \ll 1$, which allows us to use Dyson series expansion to solve Eq.~\eqref{eqn:Heisenberg_a} and only keep the lower order outcomes. Together with Eq.~\eqref{eqn:int_by_part}, we can achieve the solution for $\hat a(t)$ and $\hat a^\da(t)$ approximately with dominant terms in the following:
\begin{widetext}
\begin{equation}
\begin{split}
\hat a(t) \simeq{} \{&1 - \frac{\chi}{2\omega_c}[(\at^{*[0]})^2 e^{2i\phi_d} e^{2i\omega_c t} - \cc] + \frac{i\chi^2}{\omega_c} \int_0^t [2(|\at^\zth(t')|^2+\frac{1}{2})^2 + \frac{|\at^\zth(t')|^4}{4}+ O(\frac{\alpha^4 \omega_r}{\omega_c})] \dif t' \} \hat a \ + \\
  \Big\{& -[\frac{\chi(2|\at^\zth|^2+1)}{2\omega_c}e^{2i\phi_d} e^{2i\omega_c t} + \frac{\chi (\at^{*[0]})^2}{4\omega_c}  e^{4i\phi_d} e^{4i\omega_c t}]\\
  &+ \frac{i\chi^2}{\omega_c} \int_0^t \{[\at^\zth(t')]^2 (2|\at^\zth(t')|^2+1)+ O(\frac{\alpha^4 \omega_r}{\omega_c})\} \dif t' \Big\} \hat a^\da.
\end{split}
\end{equation}
\end{widetext}
We can see that the difference between $\hat a(t)$ and $\hat a$ is linear in $\hat a$ and $\hat a^\da$ whose coefficients are at most $O(\frac{\chi^2 \alpha^4 T}{\omega_c})$, which are far less than $1$. Let us focus on the terms without $e^{ik\omega_c t}$ factors first. Notice that the leading-order terms come from the time integration of some polynomials of $\at^\zth$ and $\at^{*[0]}$ with at most fourth order, which is denoted as $p(\at^\zth, \at^{*[0]})$ later in general. Also $\at^\zth(t) = \alpha(t) f(t)$ where $f(t)$ is a periodical function that can be written as a Fourier series with basis $e^{ik\omega_r t}$ ($k \in \mathbb{Z}$). Due to the time integration in $\frac{\chi^2}{\omega_c} \int_0^t p(\at^\zth, \at^{*[0]}) \dif t'$, only the coefficient of $e^{ik\omega_r t}$ with $k=0$ scales as $O(\frac{\chi^2 \alpha^4 T}{\omega_c}) \sim O(\frac{\chi \alpha^3}{\omega_c})$, while terms that oscillate as $e^{ik\omega_r t}$ (with nonzero $k$) only have amplitudes that scale as at most $O(\frac{\chi^2 \alpha^4}{\omega_r \omega_c})$. Therefore, for the corrections that come from $\frac{\chi}{2} (\at^\zth)^{2} ([\hat a^{\da}(t)]^2 - \hat a^{\da 2})$, they contain terms with amplitude $O(\chi \alpha^2 \cdot \frac{\chi\alpha^3}{\omega_c}) \ll O(\chi \alpha^2)$ but oscillate in the form of $f^2(t)$, whose effect is ignorable as we argued at the beginning of the section. They also contain slowly varied part with coefficients at most $O(\chi \alpha^2 \cdot \frac{\chi^2\alpha^4}{\omega_c \omega_r}) \sim O(\frac{\chi\alpha^2}{\omega_r}\cdot \frac{\chi\alpha^3}{\omega_c}\cdot\frac{1}{T})$, which are again ignorable after integration over time $T$. We can similarly demonstrate that the corrections coming from $\chi\at^\zth \at^{*[1,1]}e^{2i\phi_d} [\hat a^{\da}(t)]^2$ and other high-order terms are negligible.

Besides, after going to the reference frame with $\hat U_{\NR, 2}$, the fast-oscillating (under $\omega_c$) $\hat a^{\da 2}$ or $\hat a^\da \hat a$ terms left have amplitudes at most $O(\frac{\chi^2 \alpha^4}{\omega_c}) \sim \frac{\chi\alpha^3}{\omega_c}\cdot\frac{1}{T})$, which are sufficiently small. Finally, since we work in a small-dimensional blockade subspace with only several excitations as assumed, as well as $\frac{\chi^2 \alpha^4 T}{\omega_c} \ll 1$, the corrections due to the frame transformation $\hat U^\da_{\NR, 2}(T)$ on the final states are negligible. This concludes all the evidence that under assumptions in Eq.~\eqref{eqn:assum}, the dynamics affected by terms beyond RWA in Eq.~\eqref{eqn:H_nr} could be ignorable, in the aspect of scaling analysis.

\subsection{Both charge and flux drives}
Things will be different if we can drive both charge and flux simultaneously. In this case, our 1-photon driving term can be implemented as
\begin{equation}
\begin{split}
    &\text{Re}[\Lambda_1(t)e^{-i\omega_1(t)t}]  (\hat a + \hat a^\da) \\
    & + \text{Re}[i\Lambda_1(t)e^{-i\omega_1(t)t}] \cdot i(\hat a - \hat a^\da)\\
    ={}& \Lambda_1(t)e^{-i\omega_1(t)t} \hat a^\da + \Lambda^*_1(t)e^{i\omega_1(t)t} \hat a.
\end{split}
\end{equation}
In this way, our leading non-RWA term in Eq.~\eqref{eqn:H_nr_lab} is gone. As a result, the improved hardware controllability can also help to mitigate the non-RWA effects. 

\subsection{Rough lower bound on operation infidelity}\label{subapp:rough_l_bd}

 We want to point out that the requirement for $\frac{\chi \alpha^3}{\omega_c} \ll 1$ is actually necessary since without the assumption the $\hat H_{\NR, 2}$ term will affect the dynamics a lot. Even if we can directly implement 2-photon drive, or we can do both charge and flux drives, the requirement is still there due to the difference between the original Kerr term $\chi(\hat a + \hat a^\da)^4/12$ and the simplified one $\chi\hat a^{\da 2}\hat a^2/2$ after RWA. We can use the $\frac{\chi \alpha^3}{\omega_c} \ll 1$ condition to derive a lower bound on the infidelity of Fock $\ket{1}$ state preparation using constant $\alpha$. Since $\alpha < (\omega_c / \chi)^{1/3}$, we have
\begin{equation}\label{eq:Tmin}
    T > \frac{\pi}{2(\omega_c \chi^2)^{1/3}}.
\end{equation}
So, even in the case that power is not a constraint, and we are allowed to design $\kappa_e$ as small as possible, we still have
\begin{equation} \label{eq:errmin}
    \epsilon > c_1 \kappa_i T > \frac{3\pi (\kappa_i/\chi)}{16 (\omega_c / \chi)^{1/3}} = \frac{3\pi (\kappa_i/\chi)^{2/3}}{16 Q_i^{1/3}},
\end{equation}
where we have already put $c_1 = \frac{3}{8}$ in it, and $Q_i = \omega_c/\kappa_i$ is the internal quality factor. We denote $\epsilon_\text{min} := \frac{3\pi (\kappa_i/\chi)^{2/3}}{16 Q_i^{1/3}}$, which serves as a lower bound of the infidelity that can be achieved.

\section{Rough lower bound for input power needed}

Here we consider some extreme cases to provide a rough lower bound of the input power needed for our protocol. We assume that we could directly implement 2-photon drives so that the infidelity only comes from photon loss that $\epsilon = c_1(\kappa_e + \kappa_i) T$. Besides, we further assume that the 2-photon drives can be implemented in some power-efficient manners such that we only need to consider the power consumed by 1-photon drives. However, from Eq.~(\ref{omega1}) we still have $\Lambda_1 \simeq 2\chi\alpha^3$, and therefore the input power we need scales as $P_\inp \propto \alpha^6$. To prepare $\ket{1}$ state with constant $\alpha$, the power-dependent infidelity satisfies the following equation
\begin{equation}
    \epsilon = c_1 \frac{\pi}{2} \frac{\kappa_e + \kappa_i}{\kappa_e^{1/6}} \left(\frac{4\hbar \omega_c}{P_\inp\chi^4}\right)^{1/6}.
\end{equation}

We can choose $\frac{\kappa_e}{\kappa_i} = \frac{1}{5}$ to further minimize $\epsilon$ when $\kappa_i$ is fixed, and therefore achieve the error scaling $\epsilon\propto\kappa_i^{5/6}/(P_\inp^{1/6}\chi^{2/3})$. Equivalently, the rough lower bound of the input power $P_\inp$ for a given infidelity $\epsilon$ satisfies
\begin{equation}
    P_\inp = \frac{4(3\pi c_1)^6}{5^5}\hbar \omega_c\frac{\kappa_i^5}{\chi^4 \epsilon^6}.
\end{equation}

\begin{table*}
\caption{A summary of power needed to achieve 90\% fidelity of Fock state $\ket{1}$ with our protocol in photonic crystals designed in \cite{Choi2017SM}, as well as the check of the self-consistency requirements for RWA.}
\begin{tabular}{c|c|c|c|c|c|c|c|c|c}
\hline\hline
\makecell{Cavity\\design} & \makecell{Nonlinear\\materials} & $\omega_c / (2\pi)$ & $|\chi| / (2\pi)$ & $\kappa_i / (2\pi)$ & $\epsilon_\mi$  & $P_{\inp, 0.9}$ & $(\omega_r/\omega_c)_{0.9}$ & $(\alpha \omega_r^2/\omega_c^2)_{0.9}$ & $(\chi \alpha^3 / \omega_c)_{0.9}$\\ \hline
\multirowcell{2}{Tip} & PIC & $521$ THz & $3.5$ THz & $0.52$ GHz  & \multicolumn{5}{c}{$\chi \gg \kappa_i$} \\ \cline{2-10} & ITO & $255$ THz & $1.9$ GHz & $0.26$ GHz & $1.5 \times 10^{-3}$ & $1.3 \times 10^{-5}$ W & $5.8 \times 10^{-5}$ & $3.3 \times 10^{-9}$ & $7.1 \times 10^{-6}$  \\ \hline
\multirowcell{2}{Bridge} & GaAs & $283$ THz & $2.7$ MHz & $0.19$ GHz & $0.088$ & $4.3 \times 10^{8}$ W & $0.46$ & $112$ & $1.3$ \\ \cline{2-10} & Ge & $94.6$ THz & $9.9$ kHz & $63$ MHz & $1.76$ &  \multicolumn{4}{c}{$\epsilon_\mi > 1$}\\ \hline\hline

\end{tabular}
\end{table*}

\section{Feasibility on actual platforms}

In this section, we try to consider parameters from different kinds of experimental platforms and check if it is feasible to use our protocol to achieve some high-fidelity operations on those devices. For simplicity we still consider the Fock $\ket{1}$ state preparation task. We will check the requirements for RWA and the input power needed in order to achieve a certain fidelity derived under RWA. It is worth mentioning that, although our RWA requirements are derived under assumptions that $\alpha(t)$ is slowly varied with $\alpha(0) = \alpha(T) = 0$ and $f(t)$ itself is analytic, in this section we still focus on the method with constant $\alpha$ and $f(t)$ defined in Eq.~\eqref{SemiRot} to calculate relevant parameters. More physical $\alpha(t)$ and $f(t)$ functions will lead to a change of those $c_i$ coefficients instead of the scaling properties, so we can still use those ideal choices to get a taste of the feasibility of our protocol on different platforms.

\subsection{Optical ring resonators}

In this part, we talk about the possibility of implementing our protocol on optical ring resonators with $\chi^{(3)}$ nonlinearity. We first estimate the self-Kerr $\chi$ value with experimentally achievable platforms, and then check the RWA requirements by calculating $\epsilon_\mi$, which is a rough lower bound for infidelity achieved by perturbative analysis together with parts of the RWA requirements (See Sec.~\ref{subapp:rough_l_bd}).

The self-Kerr parameter $\chi$ in optical ring resonators with $\chi^{(3)}$ nonlinearity can be calculated by \cite{Hoff2015SM}:
\begin{equation}
    \chi = -\frac{\hbar \omega_c^2 c n_2}{2n^2 A_{\text{eff}}L},
\end{equation}
where $\omega_c$ is the frequency of the mode, $n$ and $n_2$ are the refractive index and nonlinear refractive index correspondingly, and $A_{\text{eff}}$ and $L$ are effective area and length of the waveguide. In the system proposed in Ref. \cite{Mittal2021SM}, the authors there assumed $\frac{\omega_c c n_2}{n^2 A_{\text{eff}}L} = 3.7 \times 10^{20}\ \text{W}^{-1} \text{s}^{-2}$ without pointing out which kind of materials they use nor those geometrical parameters of the resonators. Besides, the authors hope to achieve an internal loss rate $\kappa_i = 2\pi \times 25$ MHz, and they mentioned that this assumption is reasonable since the corresponding intrinsic quality factor $Q_i\sim 8\times 10^6$ is achievable with the silicon-nitride platform. This discussion indicates that they have chosen $\omega_c \approx 2\pi \times 200$ THz. With those parameters, the corresponding $\chi = -2\pi \times 3.90$ Hz. However, when putting these numbers into Eq.~\eqref{eq:errmin}, we can find out that $\epsilon_\text{min} = 102 \gg 1$, which indicates that our perturbative analysis should fall and RWA will be significantly violated if we want to use our current protocol to prepare a Fock $\ket{1}$ state.

A similar result is achieved using parameters from another literature \cite{Vernon2019SM}. In their simulation, the parameters are chosen as $\gamma_{\text{NL}} := \frac{n_2 \omega_c}{c A_{\text{eff}}} = 1\ \text{W}^{-1} \text{s}^{-2}$, together with $\omega_c = 2\pi \times 193$ THz, $n = 1.7$, and $L = 400$ \textmu m. With these numbers, we can find that $\chi = -2\pi \times 0.79$ Hz. The intrinsic Q-factor is chosen as $2 \times 10^6$, which corresponds to $\kappa_i = 2\pi \times 96.5$ MHz. This gives $\epsilon_\mi = 1151$, which again indicates that our protocol does not work.



\subsection{Photonic crystals}

Photonic crystals may perform better to increase Kerr nonlinearities with smart design by suppression of effective mode volume. Our discussion in this part builds on the work from Ref.~\cite{Choi2017SM}, where the authors discussed the $\chi$ and $\kappa$ values that can be possibly achieved using their novel design of nanocavities with ultra-small mode volumes. It is claimed to be promising for achieving a ``single-photon Kerr nonlinearity" ($\chi \sim \kappa$) regime. In that work, $\chi$ is determined by
\begin{equation}
\chi = -\frac{3\chi^{(3)}\hbar \omega^2_c}{2\epsilon_0 n^4}\frac{V_M}{V^2_\eff},
\end{equation}
where $V_\eff$ is the effective mode volume, and $V_M$ is another parameter related to cavity design with the unit of volume.

The authors there provided two types of cavity design: tip design and bridge design. With the tip cavity design there, in simulations people could achieve $\frac{QV_M}{V_\eff^2} \approx 2 \times 10^7 \lambda^{-3}$ and quality factor $Q \approx 10^6$. The $\lambda$ here is the wavelength of the mode. Besides, in that work, those authors assumed that the cavity radiation loss is much larger than material loss so that the same quality factor applies to cavities with the same design but different materials. Here we simply follow this assumption to derive $\kappa_i$.

For organic materials like J aggregate (PIC), it has $|\chi^{(3)}|/n^4 \sim 1.1 \times 10^{-15} \text{m}^2/\text{V}^2$ at $\lambda = 575$ nm, which gives $\chi = - 2\pi \times 3.5$ THz and $\kappa_i = 2\pi \times 0.52$ GHz. In this case $\chi$ is several orders of magnitude larger than $\kappa_i$, so there will be no need to use our protocol to achieve universal control. 

Inorganic materials like indium tin oxide (ITO) has $|\chi^{(3)}|/n^4 \sim 2.12\times 10^{-17} \text{m}^2/\text{V}^2$ at $\lambda = 1175$ nm. It gives $\chi = -2\pi \times 1.9$ GHz with $\kappa_i = 2\pi \times 0.26$ GHz. The parameters still lie in the strong Kerr-nonlinearity regime as $|\chi|/\kappa_i > 1$, and we only need input power $P_\inp = 1.3\times 10^{-5}$ W to achieve 90\% fidelity of a single photon state with our protocol.

Similarly, for their bridge cavity design they have  $\frac{QV_M}{V_\eff^2} \approx 5.5 \times 10^7 \lambda^{-3}$ and quality factor $Q \approx 1.5 \times 10^6$, which in general requires a small $|\chi^{(3)}|/n^4$ to achieve the same Kerr nonlinearity. But, as they mentioned, for usual semiconductor materials like gallium arsenide (GaAs) with $|\chi^{(3)}|/n^4 \sim 0.97\times 10^{-20} \text{m}^2/\text{V}^2$ at $\lambda = 1.06$ \textmu m, we can get $\chi = -2\pi \times 2.7$ MHz and $\kappa_i = 2\pi \times 0.19$ GHz. In this case, if we want to achieve $90\%$ fidelity at the optimal point we derived in Sec.~\ref{app:err_ana} under RWA, in fact we can see that $\alpha\omega_r^2/\omega_c^2 = 112 \gg 1$, which means the RWA is actually violated. If people can get a better device with $\kappa_i$ 10 times smaller (so that $\kappa_i = 2\pi \times 19$ MHz and $\kappa_i/\chi = 7.0$) while other parameters do not change, it is promising to pass the RWA requirement with $\omega_r/\omega_c = 1.5 \times 10^{-3}$, $\chi \alpha^3 / \omega_c = 1.3 \times 10^{-3}$ and $\alpha \omega_r^2/\omega_c^2 = 1.1 \times 10^{-4}$ if we want to achieve $90\%$ fidelity. The power needed will be $P = 4.3 \times 10^2$ W.
 

For bridge cavity design using Germanium (Ge) as nonlinear materials with $|\chi^{(3)}|/n^4 \sim 0.86\times 10^{-20} \text{m}^2/\text{V}^2$ at $\lambda = 3.17$ \textmu m, we have $\chi = -2\pi \times 9.9$ kHz and $\kappa_i = 2\pi \times 63$ MHz. However, we find that $\epsilon_\mi = 1.76 > 1$, which rules out the possibility of using our protocol due to the violation of RWA.



\section{Derivation of continuous version formulae}\label{app:misc}

In this section, we will use the Magnus expansion method to derive some ``continuous version" formulae including Eq.~\eqref{contiTrot} and~\eqref{eqn:coh_dev}.

We first use Magnus expansion to compute $\hat U = \mathcal{T} \exp[ -i \int_0^{\delta T} \hat H(t) \dif t]$ up to fourth order explicitly with $\hat H(t) = \hat H(\delta T - t)$. Notice that
\begin{equation}
\hat U = \exp{\sum_{m=1}^4 (-i)^m \hat \Omega_m + O[(\delta T)^5]},
\end{equation}
where
\begin{widetext}
\begin{equation}
\begin{split}
&\hat \Omega_1 = \int_0^{\delta T} \hat H_1 \dif t_1, \quad \hat \Omega_2 = \frac{1}{2} \int \dif{}^2 t \ [\hat H_1, \hat H_2], \quad \hat \Omega_3 = \frac{1}{6}\int \dif{}^3 t \ ([\hat H_1, [\hat H_2, \hat H_3]] + [\hat H_3, [\hat H_2, \hat H_1]]),\\
&\hat \Omega_4 = \frac{1}{12} \int \dif{}^4 t \ ([[[\hat H_1, \hat H_2], \hat H_3], \hat H_4] + [\hat H_1, [[\hat H_2, \hat H_3], \hat H_4]] + [\hat H_1, [\hat H_2, [\hat H_3, \hat H_4]]] + [\hat H_2, [\hat H_3, [\hat H_4, \hat H_1]]]).
\end{split}
\end{equation}
Here we use abbreviations for $\hat H_k := \hat H(t_k)$ and
\begin{equation}
\int \dif{}^k t := \idotsint\limits_{0\leq t_k \leq t_{k-1} \leq \dots \leq t_1 \leq \delta T} \dif t_1 \dots \dif t_k.
\end{equation}
By making use of the property $\hat H(t) = \hat H(\delta T - t)$, we have the following:
\begin{gather}
\hat \Omega_2 = \frac{1}{2} \int \dif{}^2 t \ [\hat H_2, \hat H_1] = -\hat \Omega_2 = 0\\
\int \dif{}^4 t \ [[[\hat H_1, \hat H_2], \hat H_3], \hat H_4] = \int \dif{}^4 t \ [[[\hat H_4, \hat H_3], \hat H_2], \hat H_1] = - \int \dif{}^4 t \ [\hat H_1, [\hat H_2, [\hat H_3, \hat H_4]]]
\end{gather}
Then, from Jacobi identity we have
\begin{equation}
\begin{split}
\int \dif{}^4 t \ [\hat H_1, [[\hat H_2, \hat H_3], \hat H_4]] &= \int \dif{}^4 t \ -[\hat H_4, [\hat H_1, [\hat H_2, \hat H_3]]] - [[\hat H_2, \hat H_3], [\hat H_4, \hat H_1]]\\
&= \int \dif{}^4 t \ -[\hat H_1, [\hat H_4, [\hat H_3, \hat H_2]]] - [[\hat H_2, \hat H_3], [\hat H_4, \hat H_1]]\\
&= \int \dif{}^4 t \ -[\hat H_1, [[\hat H_2, \hat H_3], \hat H_4]] - [[\hat H_2, \hat H_3], [\hat H_4, \hat H_1]]\\
&= -\frac{1}{2} \int \dif{}^4 t \ [[\hat H_2, \hat H_3], [\hat H_4, \hat H_1]].
\end{split}
\end{equation}
Similarly, we have
\begin{equation}
\int \dif{}^4 t \ [\hat H_2, [\hat H_3, [\hat H_4, \hat H_1]]] = -\frac{1}{2} \int \dif{}^4 t \ [[\hat H_4, \hat H_1], [\hat H_2, \hat H_3]],
\end{equation}
which will lead to $\hat \Omega_4 = 0$.
\end{widetext}

In summary, we have
\begin{equation}
\begin{split}
&\hat U = \exp{-i\hat \Omega_1 + i \hat \Omega_3 + O[(\delta T)^5]},\\
&\hat U_\tar = \exp[ -i \int_0^{\delta T} \hat H(t) \dif t] = \exp[ -i \hat \Omega_1].
\end{split}
\end{equation}
Since $\hat \Omega_3$ is in the order of $O[(\delta T)^3]$, it is easy to see that 
\begin{equation}
\hat U - \hat U_\tar = O[(\delta T)^3],
\end{equation}
which gives the proof of Eq.~\eqref{contiTrot}.

Similarly, with the Baker–Campbell–Hausdorff (BCH) formula, we can find that
\begin{equation}
\hat U_\tar^{-1/2} \hat U \hat U_\tar^{-1/2} = \exp{i \hat \Omega_3 + O[(\delta T)^5]},
\end{equation}
which gives the proof of the ``continuous version" correspondence of Eq.~\eqref{eqn:coh_dev}.

\section{Estimation of undetermined factors in Eq.~\eqref{EPSITOT}} \label{app:esti_coe}

In this section, we will show how we estimate the coefficients $c_1$, $c_2$ and $c_3$ in Eq.~\eqref{EPSITOT} when focusing on a specific task described in Sec.~\ref{app:err_ana}, which is to prepare Fock state $\ket{1}$ with constant $\alpha$ and oscillation function $f(t)$ defined in Eq.~\eqref{SemiRot}.

\subsection{Estimation of $c_1$}

In the estimation of $c_1$, we can assume that the photon blockade is perfectly achieved with ideal Hamiltonian $\hat H^\qb_\dr$ in Eq.~\eqref{eqn:h_qb_rd}, so we only need to take care of the infidelity induced by photon loss. We can start with the following Lindblad equation with operators restricted in the two-level blockade subspace:
\begin{equation}
\frac{\dif \hat \rho}{\dif t} = -i[-\chi \alpha \hat \sigma_x, \hat \rho] + \kappa \mathcal{D}[\hat \sigma_-] \hat \rho.
\end{equation}

The state is initialized as $\hat \rho(0) = \ketbra{0}{0}$. To prepare the $\ket{1}$ state, we require the evolution to last for $T = \frac{\pi}{2\chi\alpha}$. Since we work in the regime that $\kappa \ll \chi\alpha$, we will treat the loss as a perturbation. We first go to the interaction picture, and then denote $\hat U_0(t) = e^{i\chi \alpha t \hat \sigma_x}$ as well as $\hat \rho_I(t) = \hat U^\da_0(t) \hat \rho(t) \hat U_0(t)$. Therefore, the evolution of $\hat \rho_I(t)$ satisfies
\begin{equation}
\frac{\dif \hat\rho_I(t)}{\dif t} = \kappa \mathcal{D}[\hat U_0^\da(t) \hat \sigma_- \hat U_0(t)] \hat \rho_I(t).
\end{equation}
The final state $\hat \rho_I(T)$ up to first order in $\kappa$ is
\begin{equation}
\hat \rho_I(T) = \hat \rho_I(0) + \kappa\int_0^T \mathcal{D}[\hat U_0^\da(t) \hat \sigma_- \hat U_0(t)] \hat \rho_I(0) \dif t.
\end{equation}

Recall our definition for infidelity $\epsilon$ in Eq.~\eqref{infidel}, we have
\begin{equation}
\begin{split}
\epsilon &= 1 - \bra{1}\hat \rho(T)\ket{1} = 1 - \bra{0}\hat \rho_I(T)\ket{0}\\
&= -\kappa \bra{0} \left[ \int_0^T \mathcal{D}[\hat U_0^\da(t) \hat \sigma_- \hat U_0(t)] \hat \rho_I(0) \dif t \right] \ket{0}\\
&= \frac{3}{8}\kappa T,
\end{split}
\end{equation}
where $T = \frac{\pi}{2\chi\alpha}$. This result directly indicates that $c_1 = \frac{3}{8}$ in the protocol we considered.

\subsection{Estimation of $c_2$}

\begin{figure}[t!]
\centering
\includegraphics[scale = 0.49]{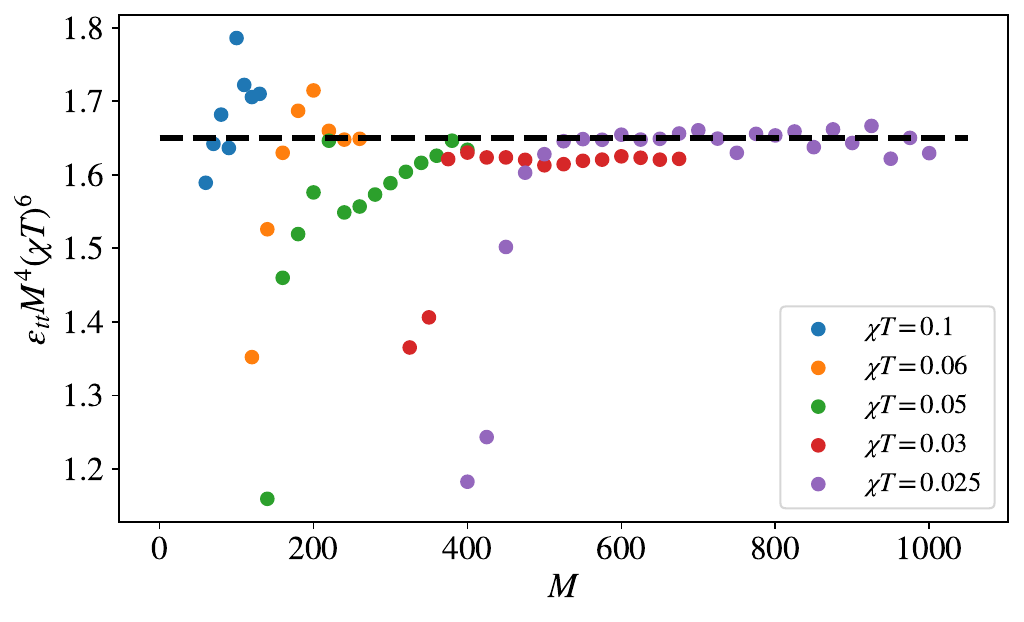}
\caption{Determine $c_2 = \epsilon_\ttt M^4 (\chi T)^6$ numerically by varying $M$ and $T$. The black dashed line is the reference for $c_2 = 1.65$.}
\label{fig:c2}
\end{figure}

Notice that $c_2$ is introduced as a factor in the expression of Trotter error $\epsilon_\ttt = c_2 / [M^4 (\chi T)^6]$, so we can numerically estimate it by calculating the infidelity of the state preparation under Hamiltonian $\hat H'_\dr[\tilde{\alpha}(t)]$ evolution while varying $M$ and $T$. We show the corresponding plot in FIG.~\ref{fig:c2} where the data points on it are chosen to satisfy $\epsilon_\ttt < 0.2$, and find out that in the small $\chi T$ (to reduce infidelity coming from photon loss) but large $M$ (to reduce $\epsilon_\ttt$) regime we have $c_2 \approx 1.65$. Since we mainly care about the scaling behavior in our protocol instead of exact numbers, we will just use $c_2 \approx 1.65$ as a rough estimation when it is needed.

\subsection{Estimation of $c_3$}

The derivation of $c_3$ is straightforward, since we just need to put the exact form of $\tilde{\alpha}(t) = \alpha f(t)$ into Eq.~\eqref{eqn:new_single_drive} and then calculating the power. As we mentioned in Sec.~\ref{app:err_ana} that $i\dot{\tilde{\alpha}}$ is the dominant term in $\Lambda_1$, and due to our specific design of $f(t)$ in Eq.~\eqref{SemiRot}, we have
\begin{equation}
|\Lambda_1| \simeq |\dot{\tilde{\alpha}}(t)| = |\alpha \dot f(t)| = \frac{\pi^2 M \alpha}{T} = \frac{\pi^3 M}{2\chi T^2},
\end{equation}
in which we have used $\alpha = \frac{\pi}{2\chi T}$. Therefore, we can finally achieve
\begin{equation}
P_\inp = \frac{|\Lambda_1|^2}{\kappa_e}\hbar \omega_c = \frac{\pi^6}{4} \hbar \omega_c \frac{M^2}{\kappa_e \chi^2 T^4},
\end{equation}
which directly gives us that $c_3 = \frac{\pi^6}{4}\hbar \omega_c$.

\section{Generation of control sequence with neural network}\label{app:num_opt}


In this part, we will discuss about the way to use neural network to generate control pulses in order to solve the imperfect blockade situation due to the assumption that we do not have the direct 2-photon drive in experiment. 

We still first go to the rotating frame and then the displaced frame, without making constraint between $\alpha(t)$ and $\Lambda_1(t)$. Besides, we can choose $\omega_1(t)$ such that there is no detuning term ($\hat a^\da \hat a$) in the new Hamiltonian. Therefore, we will end up with the following Hamiltonian:
\begin{equation}
\begin{split}
    \hat H_\dr^\new = \frac{\chi}{2}\hat a^{\da 2} \hat a^2 + \{&\chi \alpha^\new(t) \hat a^\da \hat n + \frac{\chi}{2}[\alpha^\new (t)]^2 \hat a^{\da 2}\\
    & + \Lambda_1^\new(t) \hat a^\da + \hc\}.
\end{split}
\end{equation}
In this case, both $\alpha^\new(t)$ and $\Lambda_1^\new(t)$ are controllable. We note that, although $r$ does not come into $\hat H_\dr^\new$ explicitly, we still try to achieve the effective photon blockade during evolution by optimizing the fidelity of the unitary operations or state preparation tasks within the subspace spanned by $\{\ket{0}, \dots, \ket{r}\}$ that we want to block, while adding a penalty function to penalize the average population leakage out of the blockade subspace during evolution. Specifically, we can choose $g_{\rm{u}}=\frac{1}{(r+1)T}\int_0^T \dif t \Tr[\hat U^\da(t)(\hat I - \hat \Pi_r) \hat U(t) \hat \Pi_r]$ as the penalty function in the unitary operation task, or $g_{\rm{st}}=\frac{1}{T}\int_0^T \dif t \bra{\psi(t)}(\hat I - \hat \Pi_r) \ket{\psi(t)}$ for the state preparation task, where as in the main text $\hat \Pi_r = \sum_{n=0}^r \ketbra{n}$ is the projection operator onto the blockade subspace. Here we can understand the dynamics as that, the states that initially lie in the blockade subspace will always evolves within it with small leakage, by keeping track of the evolution with a proper time-dependent displace frame $\alpha^\new(t)$. This also provides another advantage that we do not have to take a large dimension cutoff which always associates with numerical computation in bosonic systems, since we just have to focus on the dynamics confined in a small subspace with slight leakage.


To encode the variables $\alpha^{\rm{new}}(t)$ and $\Lambda_1^{\rm{new}}(t)$ in the optimization problem, we use sinusoidal representation neural network with three layers that takes $t$ as an input and returns the real and imaginary part of the corresponding control parameter by $\Phi^{(t)}=\mathbf{W}_2 \sin(\mathbf{W}_1 \sin(\mathbf{W}_0 t +\mathbf{b}_0) + \mathbf{b}_1)+\mathbf{b}_2$, where $\mathbf{W}_0 \in \mathbb{R}^{20\times 1}$, $\mathbf{W}_1 \in \mathbb{R}^{20\times 20}$, and $\mathbf{W}_2 \in \mathbb{R}^{2\times 20}$, and $\mathbf{b}_i$ are vectors with a dimension consistent with their corresponding $\mathbf{W}_i$~\cite{sitzmann2020implicitSM}. We then maximize $F_{\rm{task}}(\boldsymbol{\alpha}) - g(\boldsymbol{\alpha})$ by finding an appropriate set of $\mathbf{W}_i$ and $\mathbf{b}_i$ that specifies the controls, where $F_{\rm{task}}(\boldsymbol{\alpha}) = \abs{\Tr[\hat U_\tar^\da \hat U(T)]}^2/(r+1)^2$ for unitary operations and $F_{\rm{task}}(\boldsymbol{\alpha}) = \abs{\braket{\psi_\tar}{\psi(T)}}^2$ for state preparation tasks. To do this we first fix the total evolution time $\chi T$ and perform gradient based optimization until a convergence criteria is satisfied. After finding $\alpha^\new(t)$ and $\Lambda_1^\new(t)$ at the end of the optimization, we compute the infidelity by taking the photon loss $(\kappa_i + \kappa_e)\mathcal{D}[\hat a]$ into account and also the input power needed. We fix $\kappa_i$ and vary $\kappa_e$ to find infidelity as a function of the required power.

Note that as opposed to the Trotter based method with the sine-basis ansatz of Eq.~\eqref{alpha_t}, we do not impose $\alpha(0)=\alpha(T)=0$ for the neural network ansatz to allow for more flexibility. These potential non-zero values in the beginning and end of the protocol can be realized by fast displacements. It is also possible to impose a constraint on these values by adding another penalty term to the cost function. 

\begin{figure}[t!]
\centering
\includegraphics[scale = 0.42]{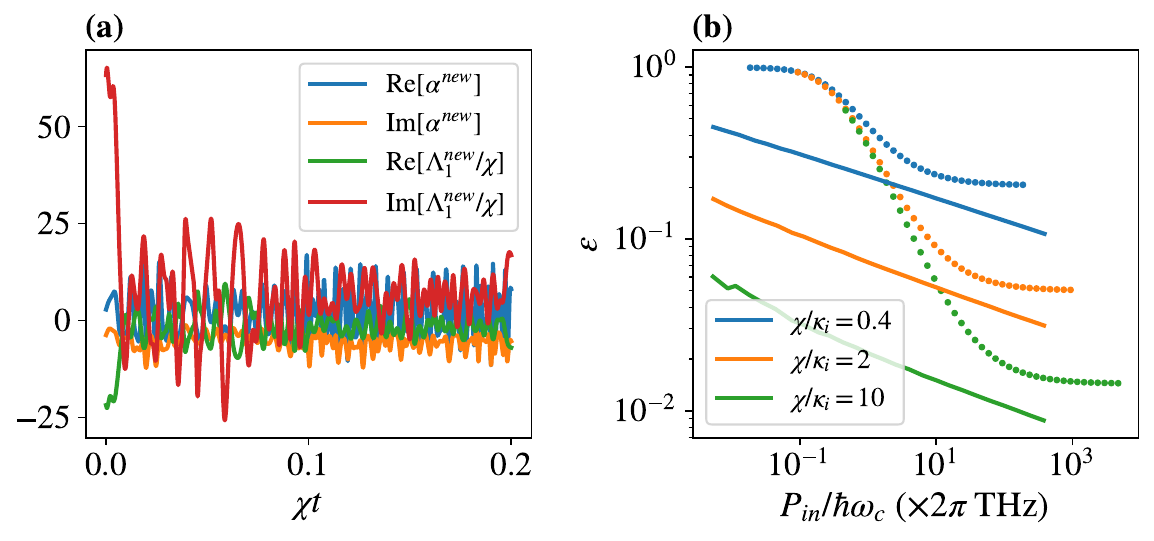}
\caption{(a) The outcome of pulse sequences $\alpha^\new(t)$ and $\Lambda_1^\new(t)$ from the neural network with fixed $\chi T = 0.2$. (b) The comparison of maximum power needed and infidelities of state preparation between pulses from neural network optimization (shown with dots) with fixed total time $T$ and Trotter-based design (shown with solid lines) with optimized $T$ and $M$. }
\label{fig:NN}
\end{figure}

To compare this method with the Trotter scheme, we consider the single photon Fock $\ket{1}$ state preparation task. Unlike the protocol examined in Sec.~\ref{app:err_ana}, where the total time $T$ can be optimized analytically based on $\chi$, $\kappa_i$, and the input power allowed, here we fix $T$ without any optimization attempt, and vary $\kappa_i$ to compare the infidelity as a function of the power for the pulse sequence generated by the neural network with those obtained in Sec.~\ref{app:err_ana} based on analytical intuition. We show an example of the outcome in FIG.~\ref{fig:NN}.

We believe that the inferior performance of the neural network is due to the complexity of finding highly oscillatory functions (as also seen in the Trotter based scheme) required for relaxing the $\Lambda_2(t)$ requirement. Solving ordinary differential equations with highly oscillatory functions is computationally costly and increases the optimization time. Therefore, even though, in principle, a highly expressive neural network is capable of representing the solutions found using the Trotter scheme, it is difficult to find those (or better) solutions directly using our optimization techniques. It is interesting to devise more efficient optimization methods using prior information from the success of the Trotter based scheme. We leave these questions for future work. 

\bibliography{ref_SM}